\documentclass[12pt]{article}
\usepackage{amsmath}
\usepackage{graphicx}
\usepackage{enumerate}
\usepackage{natbib}
\usepackage{setspace,graphicx,epstopdf,amsfonts,amssymb,amsthm}
\usepackage{mathtools}
\usepackage[colorlinks,citecolor=blue,urlcolor=blue,filecolor=blue,backref=page]{hyperref}
\usepackage[table,xcdraw]{xcolor}
\usepackage{colortbl}
\usepackage{filecontents}
\usepackage{multirow}
\usepackage{caption}
\usepackage{subcaption}
\usepackage{bm}
\usepackage{hyphenat}
\newcommand{\blind}{1}

\begin{document}

\def\spacingset#1{\renewcommand{\baselinestretch}%
{#1}\small\normalsize} \spacingset{1}

%%%%%%%%%%%%%%%%%%%%%%%%%%%%%%%%%%%%%%%%%%%%%%%%%%%%%%%%%%%%%%%%%%%%%%%%%%%%%%

\if1\blind
{
  \title{\bf Bayesian adaptive and interpretable functional regression for exposure profiles}
  \author{Yunan Gao\thanks{PhD student, Department of Statistics, Rice University (\href{mailto:yunan.gao@rice.edu}{yunan.gao@rice.edu}).} \ and 
   Daniel R.  Kowal\thanks{Dobelman Family Assistant Professor, Department of Statistics, Rice University (\href{mailto:Daniel.Kowal@rice.edu}{daniel.kowal@rice.edu}). Research was sponsored by the Army Research Office (W911NF-20-1-0184), the National Institute of Environmental Health Sciences of the National Institutes of Health (R01ES028819), and the National Science Foundation (SES-2214726). The content, views, and conclusions  contained in this document are those of the authors and should not be interpreted as representing the official policies, either expressed or implied, of the Army Research Office, the North Carolina Department of Health and Human Services, Division of Public Health, the National Institutes of Health, or the U.S. Government. The U.S. Government is authorized to reproduce and distribute reprints for Government purposes notwithstanding any copyright notation herein.}}
  \maketitle
} \fi
%Research was sponsored by the Army Research Office (W911NF-20-1-0184) and the National Institute of Environmental Health Sciences of the National Institutes of Health (R01ES028819). The content, views, and conclusions  contained in this document are those of the authors and should not be interpreted as representing the official policies, either expressed or implied, of the Army Research Office, the North Carolina Department of Health and Human Services, Division of Public Health, the National Institutes of Health, or the U.S. Government. The U.S. Government is authorized to reproduce and distribute reprints for Government purposes notwithstanding any copyright notation herein.
\if0\blind
{
  \bigskip
  \bigskip
  \bigskip
  \begin{center}
    {\LARGE\bf Bayesian adaptive and interpretable functional regression for exposure profiles}
\end{center}
  \medskip
} \fi

\bigskip
\begin{abstract}
Pollutant exposure during gestation is a known and adverse factor for birth and health outcomes. However, the links between prenatal air pollution exposures and educational outcomes are less clear, in particular the critical windows of susceptibility during pregnancy. Using a large cohort of students in North Carolina, we study the link between prenatal daily $\mbox{PM}_{2.5}$ exposure and 4th end-of-grade reading scores. We develop and apply a locally adaptive and highly scalable Bayesian regression model for scalar responses with functional and scalar predictors. The proposed model pairs a B-spline basis expansion with dynamic shrinkage priors to capture both smooth and rapidly-changing features in the regression surface. The model is accompanied by a new decision analysis approach for functional regression that extracts the critical windows of susceptibility and guides the model interpretations. These tools help to identify and address broad limitations with the  interpretability of functional regression models. Simulation studies demonstrate more accurate point estimation, more precise uncertainty quantification, and far superior window selection than existing approaches. Leveraging the proposed modeling, computational, and decision analysis framework, we conclude that prenatal $\mbox{PM}_{2.5}$ exposure during early and late pregnancy is most adverse for 4th end-of-grade reading scores. 
\end{abstract}

\noindent%
{\it Keywords:}  decision analysis, functional data analysis, nonparameteric regression, shrinkage, spline 
\vfill

\newpage
\spacingset{1.8} % DON'T change the spacing!
\section{Introduction}

% Emphasize interpretability AND selecting critical windows 
% Where to place the lit review? Shorten?

%\subsection{Prenatal $\mbox{PM}_{2.5}$ exposures and educational outcomes}

Prenatal exposure to air pollution is related to a wide range of adverse birth, health, and behavioral outcomes in children. These outcomes include smaller fetal growth measurements \citep{leung2022exposure},  low birth weight \citep{vsram2005ambient, kloog2012using}, infant negative affectivity \citep{rahman2021prenatal}, and childhood asthma \citep{hazlehurst2021maternal, leon2015prenatal}, among many others. However, the link between air pollution exposure during pregnancy and later educational outcomes is less clear. Previous studies have demonstrated the adverse effects of prenatal air pollution exposure on 
neuropsychological development  \citep{suades2015air}, brain structure \citep{guxens2018air}, memory function and attention \citep{chiu2016prenatal}, and  autism diagnosis \citep{kalkbrenner2015particulate}. Although these studies suggest that a link between prenatal air pollution exposure and adverse educational outcomes is plausible, they     do not directly consider educational outcomes and   do not estimate the critical windows of susceptibility during pregnancy.

%This link might be expected, not only because of the way adverse health experiences negatively impact educational outcomes, but also because of the negative impact of air pollution during fetal life on child neuropsychological development  \citep{suades2015air}, as well as the association between prenatal air pollution exposure and brain structural alterations of the cerebral cortex \citep{guxens2018air}. Such impacts at early ages could have significant long-term consequences. For example, \cite{chiu2016prenatal} has shown that prenatal $\mbox{PM}_{2.5}$ exposure is associated with poorer function of memory and attention, and \cite{kalkbrenner2015particulate} has shown the association between increasing odds of autism diagnosis and $\mbox{PM}_{10}$ exposure during pregnancy. Therefore, the aim of this application study is to examine the association of prenatal exposure to air pollutes and educational outcomes with adjustment for several socioecomomic and demographics variables. 

To address these limitations, we study daily $\mbox{PM}_{2.5}$ exposure during gestation and its impact on 4th end-of-grade (EOG) standardized testing for a large cohort of students in North Carolina (NC). The dataset is created by linking multiple administrative datasets in NC that include birth and demographic information, blood lead level measurements, socio-economic status, and 4th EOG reading test scores on $n=98,159$ mother-child pairs; see Table~\ref{tab:variable} for the primary variables and  \cite{bravo2021effects} and \cite{FelKow2022}  for additional details. Daily $\mbox{PM}_{2.5}$ exposure is computed at the mother's home address using Fused Air Quality Surface using Downscaling  (FAQSD)  data provided by the United States Environmental Protection Agency. The goal of our analysis is to estimate and characterize the effects of prenatal exposure to $\mbox{PM}_{2.5}$ on educational outcomes, and in particular to identify the time periods during gestation---if any---that are  predictive of adverse educational outcomes, while adjusting for important confounding variables.

For this task, our  primary modeling tool is \emph{scalar-on-function regression} (SOFR): 
\begin{equation}
    y_i =  \mu + \bm z_i'\bm\alpha +  \int_{\mathcal{T}_i} X_i( t)\beta (t)\ dt + \epsilon_i, \quad \epsilon_i \stackrel{iid}{\sim} \mathcal{N}(0, \sigma^2), \quad  i=1,...,n \label{SOFR}
\end{equation}
which links the  standardized 4th EOG reading score $y_i \in \mathbb{R}$ with daily $\mbox{PM}_{2.5}$ exposure during gestation $X_i\!:\mathcal{T}_i \rightarrow \mathbb{R}$ and other scalar covariates $\bm z_i \in \mathbb{R}^p$  (see Table~\ref{tab:variable}) for each mother-child pair $i=1,\ldots,n$.
The crucial term is $ \int_{\mathcal{T}_i} X_i( t)\beta (t)\ dt$, which represents the cumulative effect of daily exposure to $\mbox{PM}_{2.5}$ during the gestational period $\mathcal{T}_i$ on 4th EOG reading scores (adjusting for $\bm z_i$). The compact domains $\mathcal{T}_i \subseteq \mathcal{T} \subset \mathbb{R}$ are specific to each mother-child pair to allow for varying gestational periods (see Section~\ref{apps}).

%For simplicity, our methods development will often omit the scalar covariates $\bm z_i$. 
 %This type of problem is particularly challenging because the trajectories are usually high dimensional, highly correlated, and often collected on irregularly-spaced locations over the domain. 

\begin{table}
\centering 
\caption{Variables in the NC dataset. Data are restricted to individuals with 30-42 weeks of gestation, 0-104 weeks of age-within-cohort, mother's age 15-44, \texttt{Blood\_Lead} $\leq$ 10, birth order $\leq$ 4, no current limited English proficiency, and residence in NC at the time of birth and the time of 4th EOG test.}\label{tab:variable}
\begin{tabular}{|l|l|}
\hline 
\multicolumn{2}{|l|}{\cellcolor[HTML]{C0C0C0}\textbf{Air quality during gestation}}                                                                                                                                                                          \\ \hline
\texttt{Prenatal\_$\mbox{PM}_{2.5}$}   & \begin{tabular}[c]{@{}l@{}} Daily $\mbox{PM}_{2.5}$ level estimated at the 2010 census tract of the  \\mother's 
home address  (with length  \texttt{Gestation})
\end{tabular}                                                                               \\ \hline
\multicolumn{2}{|l|}{\cellcolor[HTML]{C0C0C0}\textbf{Birth information}}                                                                                                                                                                                     \\ \hline
%\texttt{LBW}          & Low Birth Weight ($< 2500$ mg)? (1 = Yes) \\ \hline
\texttt{mEdu}         & \begin{tabular}[c]{@{}l@{}}Mother's education group at the time of birth \\(NoHS = no high school diploma, HS = high school diploma, \\higherHS = some college/associates or higher) \end{tabular} \\ \hline
\texttt{mRace}        & \begin{tabular}[c]{@{}l@{}}Mother's race/ethnicity group (Non-Hispanic (NH) White, \\ NH Black, Hispanic)\end{tabular}                                                                                         \\ \hline
\texttt{mAge}         & Mother's age at the time of birth                                                                                                                                                                                                    \\ \hline
\texttt{Male}         & Male infant? (1 = Yes)                                                                                                                                                                                                               \\ \hline
\texttt{Smoker}       & Mother smoked? (1 = Yes)                                                                                                                                                                                                             \\ \hline
\texttt{Gestation} & Clinical estimate of the gestation length (days)
\\ \hline
\texttt{BirthMonth} & Birth month of the student 
\\ \hline
\multicolumn{2}{|l|}{\cellcolor[HTML]{C0C0C0}\textbf{Education/End-of-grade (EOG) test information}}                                                                                                                                                                       \\ \hline

\texttt{Reading\_Score} & \begin{tabular}[c]{@{}l@{}} Standardized score for the (chronologically first)  4th EOG\\  reading test  \end{tabular} \\ \hline

\texttt{Age\_w\_cohort}  & \begin{tabular}[c]{@{}l@{}} Age-within-cohort: 
the relative age of each student within their \\cohort (see the supplementary material for details and summary \\statistics)
\end{tabular} 
\\ \hline
                                                                                                                                    
%\multicolumn{2}{|l|}{\cellcolor[HTML]{C0C0C0}\textbf{Blood lead level information}}                                                                                                                                                                          \\ \hline
%\texttt{Blood\_lead}  & Blood Lead Level (micrograms per deciliter)                                                                                                                                                                                          \\ \hline
\multicolumn{2}{|l|}{\cellcolor[HTML]{C0C0C0}\textbf{Blood lead surveillance}}                                                                                                                                                                                     \\ \hline
\texttt{Blood\_lead}  & Blood lead level (micrograms per deciliter) \\ \hline

\multicolumn{2}{|l|}{\cellcolor[HTML]{C0C0C0}\textbf{Social/Economic status}}                                                                                                                                                                                     \\ \hline

\texttt{EconDisadvantage}         & \begin{tabular}[c]{@{}l@{}} Economically disadvantaged students are indicated by \\participation in the free/reduced 
price lunch program \\(1 = Participation in the program) \end{tabular} \\ \hline

\end{tabular}

\end{table}

SOFR is broadly useful for the medical and behavioral sciences, which often involve mapping the relationship between a scalar response and data collected repeatedly along some continuous domain (such as time-within-gestation) \citep{morris2015functional}. Because the functional covariates $X_i$ are usually high dimensional, highly correlated, and often collected on irregularly-spaced locations over the domain, regularization of $\beta$ is a central focus in SOFR. Regularization is enforced via penalties or priors to guard against both  overfitting $\beta$ and   the multicollinearities induced by the within-function correlations of $\{X_i\}$. Classical approaches expand $\beta$ using a known basis expansion and introduce a prior or penalty that encourages smoothness, such as splines with penalties on the differenced coefficients \citep{marx1999generalized,james2002generalized} or wavelets with sparsity priors or penalties \citep{brown1998multivariate, morris2006wavelet, morris2008bayesian}. %, but wavelets require a dyadic number of equally-spaced observations of each $X_i$.
Principal components analysis can be applied directly to \eqref{SOFR}, but does not account for the ordering within the $X_i$ curves \citep{cardot1999functional,muller2005generalized}. % For a thorough review of existing functional regression models, see \cite{}.

For estimating the effects of cumulative exposures, the distributed lag model (DLM) is a widely-used variant of \eqref{SOFR} that replaces the integral with a multiple regression equation featuring lagged exposure measurements \citep{schwartz2000distributed}. As in SOFR, the DLM emphasizes flexible modeling of the regression surface along with regularization of the coefficients, and has been generalized for spatio-temporal data \citep{warren2012spatial} and tree-based regression models  \citep{mork2022treed}. Despite the similarities between SOFR and DLMs, we prefer the representation in \eqref{SOFR} because it does not require the exposures $X_i(t)$ to be observed at the same time points for all subjects. In particular,  gestational length ranges from 30 to 42 weeks in our dataset, which requires careful consideration of the subject-specific domain $\mathcal{T}_i$ in \eqref{SOFR}. Nonetheless, the proposed modeling, computational, and decision analysis strategies remain relevant for DLMs.

This paper highlights and addresses two fundamental and significant challenges for SOFR (and thus DLMs). 
First,  the performance and utility of model \eqref{SOFR} hinges on the ability to estimate $\beta$. If the model for $\beta$ fails to capture the shape of the true regression function---which may vary smoothly or exhibit rapid changes---then  the estimates of $\beta$ will be biased, the uncertainty quantification for $\beta$ will be poorly calibrated, and the predictions of $y$ will be suboptimal. Existing methods for SOFR commonly produce interval estimates for $\beta$ that are far too conservative, which limits the power to detect important covariate effects (see Section~\ref{sims}). Thus, it is critical to produce estimation and inference tools for $\beta$ that adapt to both smooth and rapid changes and provide precise yet well-calibrated uncertainty quantification. At the same time, computational scalability is essential: our dataset contains $n\approx 100,000$ mother-child pairs and hundreds of observation points for each $X_i$.   Our modeling and computing strategies emphasize both \emph{adaptability} and \emph{scalability}. 

The second and more subtle challenge is that of interpretability: given an estimate and inference of the regression coefficient function $\beta$, how does one interpret the results? More concretely, consider $\beta(t^*)$ at a specific point $t^* \in  \mathcal{T}$. In the context of \eqref{SOFR}, we may be tempted to interpret the coefficient function as  
\begin{align}
\label{n-interp-1}
\beta(t^*) &\approx \int_{N(t^*)} \{X( t) + 1\}\beta (t)\ dt - \int_{N(t^*)} X( t) \beta (t)\ dt \\
\label{n-interp-2}
& = \mathbb{E}\left[y \mid \bm z_i, \{X(t) + 1\}_{t \in N(t^*)}\right] -  \mathbb{E}\left[y \mid  \bm z_i,\{X(t)\}_{t \in N(t^*)}\right]
\end{align}
where $N(t^*)$ is a small neighborhood around $t^*$ and the expectations also condition on the parameters  ($\mu,\bm \alpha, \beta$). Informally, \eqref{n-interp-1}--\eqref{n-interp-2} suggests that the regression function $\beta$ at time $t^*$ corresponds to the change in the expected response variable for a one-unit increase of $X$ in a neighborhood of $t^*$, \emph{all else equal}.  Yet for functional covariates, this latter qualification is usually not meaningful: given a trajectory $\{X(t)\}_{t \in\mathcal{T}}$, it is difficult to envision that same trajectory, but with $X(t)$ replaced  by $X(t)+1$ \emph{only in a small neighborhood of $t^*$}. Such an abrupt and localized perturbation of the trajectory is typically not consistent with the  data-generating process, especially when the curves $X_i$ are modeled as smooth functions. 
These difficulties propagate more broadly, including effect  directions
and selection of critical windows of susceptibility. In particular, the interpretation of $\beta(t) > 0$ for  $t \in \mathcal{T}^+$ and $\beta(t) < 0$ for  $t \in \mathcal{T}^- $ for subdomains $\mathcal{T}^+, \mathcal{T}^-\subset \mathcal{T}$ is nontrivial, especially when the curves $X_i$ exhibit structured (e.g., seasonal) correlations; this issue is discussed in detail  in Section~\ref{apps}. 
Similar warnings regarding interpretability were issued by \cite{dziak2019scalar}, although they did not suggest general purpose tools to resolve these challenges.  As such, we are motivated to produce \emph{model summarization} techniques that enable both interpretable estimation and powerful window selection for SOFR.

Our main methodological contribution is a new Bayesian adaptive scalar-on-function regression (BASOFR) model  paired with a decision analysis strategy to select critical windows of susceptibility and deliver more interpretable model summaries. 
%\subsection{Overview of the proposed approach} 
%The statistical methodology to analyze these data must be able to (i) adapt to smooth or rapid changes in the coefficient function $\beta$, (ii) scale  to the order of $n\approx 100,000$ observations and about  300 observation points  per function, and (iii) provide the inferential tools to describe, interpret, and visualize the model-based results. 
The BASOFR specifies a B-spline basis expansion for $\beta$ and a dynamic shrinkage prior \citep{kowal2019dynamic} on the (second differenced) basis coefficients. Crucially, this local and adaptive shrinkage prior encourages smoothness yet can capture rapid changes in $\beta$, which 
produces better point estimates and more precise uncertainty quantification, especially in the presence of both smooth and rapidly-changing features (see Section~\ref{sims}).  Importantly, the proposed modeling structure admits a highly scalable Gibbs sampling algorithm, which is necessary for our data analysis with $n \approx 100,000$ and hundreds of observation points per curve $X_i$. %This fully Bayesian model produces posterior and predictive uncertainty quantification without any cross-validation. 

%Using a B-spline expansion for $\beta$, we construct an aggressive shrinkage prior on the second differences of the basis coefficients. Crucially, unlike existing strategies that impose constant regularization on these increments (e.g., \citealp{marx1999generalized}), we apply a shrinkage prior that is both \emph{local}  and \emph{adaptive}. In particular, we deploy the \emph{dynamic horseshoe prior} \citep{kowal2019dynamic}, which exhibits horseshoe-like shrinkage \citep{carvalho2010horseshoe} behavior with dynamic dependence over the domain $\mathcal{T}$. The near-sparsity of this prior encourages $\beta'' \approx 0$, which implies local linearity of $\beta$, while the heavy tails of the prior allow large increments in the slope and therefore rapid changes in $\beta$. By adopting the  \emph{dynamic} horseshoe prior, we obtain more adaptive regularization, which produces better point estimates and more precise uncertainty quantification, especially in the presence of both smooth and rapidly-changing features. Importantly, the proposed modeling structure---including both the basis expansion and  the dynamic horseshoe representation---admits a convenient and highly scalable Gibbs sampling algorithm. This fully Bayesian model produces posterior and predictive uncertainty quantification without any tuning parameters or cross-validation. 

Leveraging the BASOFR output, we develop a decision analysis approach to select the critical windows of $\mathcal{T}$ and provide interpretable model summaries. A crucial observation is that estimation and uncertainty quantification for $\beta$ is \emph{not} sufficient for selecting critical windows of susceptibility: some  decision analysis or other selection criteria (see below) are required. We propose to extract locally constant   point estimates from the BASOFR model---or more generally, any Bayesian SOFR model---which feature estimates of the form $\hat \beta(t) = \hat \delta_k$  for $t \in \mathcal{T}_k$  and $\{\mathcal{T}_k\}$ a learned partition of $\mathcal{T}$. In conjunction, the estimated coefficients and partition $\{\hat \delta_k, \mathcal{T}_k\}$ identify effect sizes, effect directions, and  critical windows of susceptibility. These locally constant estimates also provide a partial resolution to the challenges raised by \eqref{n-interp-1}--\eqref{n-interp-2}: namely, $\hat \delta_k$ estimates the change in the expected response variable for a one-unit increase in the aggregated trajectory $X(\mathcal{T}_k) \coloneqq \int_{\mathcal{T}_k} X(t) \ dt$ while holding $\{X(\mathcal{T}_j)\}_{j  \ne k}$ constant. Here, the  notion of \emph{all else equal} is  more plausible and  less restrictive: it refers to distinct regions of the domain---rather than neighboring time points---and only requires the aggregated trajectories $\{X(\mathcal{T}_j)\}_{j  \ne k}$---rather than the entire $\{X(t)\}_{t \in \mathcal{T}}$ paths---to be held constant outside of $\mathcal{T}_k$. %These locally constant estimates provide direct information regarding  the regions of  $\mathcal{T}$ that are most important for predicting  the response, and can be used to reduce storage costs for the functional trajectories. %Because we implement a more general and adaptive  Bayesian SOFR model, we can use this model to evaluate and compare the estimators across distinct partitions $\{\mathcal{T}_k\}$. By varying the resolution or complexity of the partition, we are able to search for and identify the simplest structures that match or nearly match the predictive accuracy of the full and unrestricted model. 
These tools contribute minimal computational cost and complement more traditional posterior summaries, such as expectations and credible intervals.

%, directly assist in SOFR model interpretability, and are compatible with any Bayesian SOFR model. 

The proposed decision analysis approach deviates from the vast majority of methods for critical window selection, which rely on pointwise credible intervals for $\beta$ \citep{warren2012spatial,leon2015prenatal,wilson2017bayesian,bose2017prenatal,lee2018prenatal} or other marginal criteria  \citep{warren2020critical}  under a Bayesian SOFR or DLM. However, it has been shown that \emph{variable} selection based on credible intervals---i.e., variables are selected if the credible intervals exclude zero---is severely underpowered and overconservative, especially compared to recent decision analysis strategies \citep{kowal2020bayesianFOSR,kowal2022bayesian,kowal2021subset}. We confirm this effect for \emph{window} selection (Section~\ref{sims_2}), which implicitly warns that popular existing methods may be erroneously omitting key windows of susceptibility.

Our decision analysis approach continues a line of research on posterior summarization of Bayesian models, which has been directed primarily for variable selection, including   linear regression  \citep{hahn2015decoupling}, graphical models \citep{bashir2019post},  seemingly-unrelated regressions \citep{puelz2017variable}, and  function-on-scalars regression \citep{kowal2020bayesianFOSR}. The window selection problem is more closely related  to change point detection than variable selection, but specific to the regression coefficient function $\beta$ in \eqref{SOFR} rather than  observed data.  Related, the frequentist approach of \cite{james2009functional}  imposes  sparsity on the derivatives of $\beta$, but requires specification of several tuning parameters and does not provide uncertainty quantification for $\beta$.  
%Hence, our goal is  to identify a small set of regions $\{\mathcal{T}_k\}$ for which the corresponding locally constant effects $\hat \delta_k$ nearly match the predictive ability of the full and unrestricted SOFR model.

An intuitive and alternative Bayesian approach is to place a prior on $\beta$ that restricts the regression function to be locally constant with unknown levels and partitions. This strategy, called BLISS \citep{grollemund2019bayesian}, faces a substantial computational burden and does not scale to moderate or large datasets such as ours (see Figure~\ref{fig:comptime_vs}). In addition, BLISS requires careful specification of multiple hyperparameters, including a fixed choice for the number of partitions. \cite{grollemund2019bayesian} proposed to  fit separate models for each specified number of partitions and then compare them using BIC. Naturally, this exercise further increases the computational cost. Regardless, we emphasize that the prior alone cannot \emph{select} windows: whether using BASOFR, BLISS, or any other Bayesian SOFR model, some decision analysis or selection criterion is still required. Our two step procedure---fitting the scalable and adaptive BASOFR model and summarizing the output  using customized decision analysis---circumvents these computational challenges yet still provides posterior uncertainty quantification, locally constant point estimates, and powerful window selection.

This article is outlined as followed. Section~\ref{model_method} presents the proposed BASOFR model.  Section~\ref{PS} contains the decision analysis approach for window selection and model summarization. Section~\ref{sims} contains a detailed simulation study. The  methods are applied to the NC data  in Section~\ref{apps}. Section~\ref{conclude} concludes. Online supplementary material includes computational details, additional simulation results, and supporting information about the NC data. \texttt{R} code is available at \url{http://github.com/YunanGao/BASOFR}.

\section{Bayesian Adaptive Scalar-on-Function Regression}\label{model_method}
The core task in fitting the SOFR model \eqref{SOFR} is to learn the unknown regression coefficient function $\beta$ given observations $\{(X_i, y_i)\}_{i=1}^n$; here we omit the scalar covariates $\bm z_i$ for simplicity. With real data, the functions are observed at discrete points: $\bm x_i = (X_i(t_{i,1}),\ldots, X_i(t_{i, m_i}))'$, where the $m_i$ observation points may be unequally-spaced or differ from subject to subject. Although it is tempting to apply (non-functional) linear regression models to $\{(\bm x_i, y_i)\}$, there are several drawbacks to this approach. First, the resulting model will be high dimensional with highly correlated covariates, and therefore requires regularization. Second, such a model fails to account for the ordering among the covariates with respect to the domain $t_{i,j} \in \mathcal{T}$, which is crucial information for both regularization and interpretation. Lastly, it is unclear how to apply this approach when the functional covariates are not observed on a common grid, which occurs for our application (Section~\ref{apps}). 

We instead pursue a basis expansion strategy for both the functional predictors and the regression coefficient function \citep{ramsay2005functional}. By expanding $X_i(t) = \sum_{k=1}^{K_X}X_{ik}^* \phi_k(t)$ and $ \beta(t) = \sum_{k=1}^{K_B}B_k^* \psi_k(t)$ for known basis functions $\{\phi_k(\cdot)\}_{k=1}^{K_X}$ and $\{\psi_k(\cdot)\}_{k=1}^{K_B}$ and  unknown coefficients $\{X_{ik}^*\}_{k=1}^{K_X}$ and $\{B_k^*\}_{k=1}^{K_B}$, the key term in \eqref{SOFR} simplifies to 
\begin{equation}
   \int_{\mathcal{T}_i}X_i(t)\beta(t)\ dt = \mathbf{X}_i^*\mathbf{J}_i^{\phi,\psi}\bm B^* = \mathbf{X}_i^{**}\bm B^*  \label{mlr}
\end{equation}
where  $\mathbf{X}_i^* = (X_{i1}^*,...,X_{iK_X}^*), \bm B^* = (B_1^*,...,B_{K_B}^*)$, and $\mathbf{J}_i^{\phi,\psi} = [\int_{\mathcal{T}_i}\phi_j(t)\psi_k(t)\ dt]_{jk}$. The basis expansions resolve the difficulties with unequally-spaced or non-common observation points for the functional predictors $\{X_i\}$, since we instead work  with the coefficients $\bm X_i^*$. In addition, the basis expansion of each $X_i$ serves as a regularization tool to smooth over the measurement errors associated with the direct observations $\bm x_i$. Lastly, the representation in \eqref{mlr} shows that fitting the SOFR \eqref{SOFR} can be made equivalent to fitting a multiple linear regression model with covariates $\mathbf{X}_i^{**}=\mathbf{X}_i^*\mathbf{J}_i^{\phi,\psi}$, which is known. Additional details regarding the basis expansions are provided in the supplementary material.

Under these basis expansions, estimation and inference on the basis coefficients $\bm B^*$ is sufficient for estimation and inference on the regression function $\beta$. Thus, a prior on $\bm B^*$ implies a prior on $\beta$. Yet despite the promise of the multiple linear regression interpretation of \eqref{mlr}, the prior on $\bm B^*$ must be specified carefully. Common shrinkage priors for regression are designed to shrink redundant linear coefficients to zero. However,  sparsity in $\bm B^*$ does not guarantee smoothness or other desirable properties of $\beta$. More specifically, the choice of prior on $\bm B^*$ cannot be decoupled from the choice of basis. For instance, when $\bm B^*$ is assigned a Gaussian prior, the resulting coefficient function  is a Gaussian process with covariance function $\mbox{Cov}\{\beta(t), \beta(s)\} = \sum_{k,\ell} \psi_k(t) \psi_\ell(s) \mbox{Cov}(B_k^*, B_\ell^*)$. Thus, the prior on $\beta$ inherits key properties from both the basis functions and the prior on  $\bm B^*$.

\begin{figure}
    \centering
    \includegraphics[width=.8\textwidth]{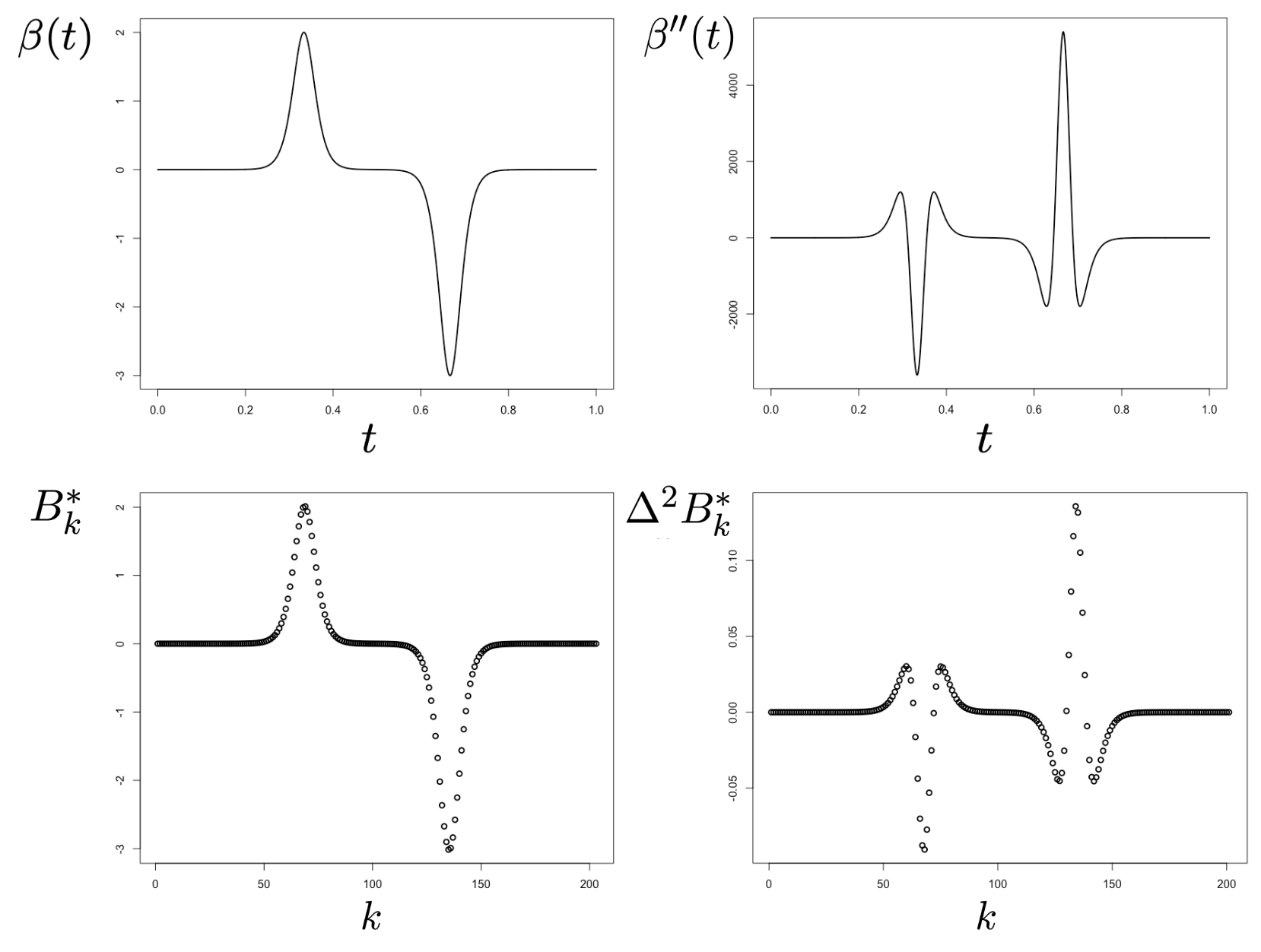}
\caption{The nonlinear function $\beta(t)=8\{2+\exp(20-60x)+\exp(60t-20)\}^{-1} - 12 \{2 + \exp(40-60t)+\exp(60t-40)\}^{-1}$ ({\bf top left}); the second derivatives $\beta''$  ({\bf top right});  the  (ordinary least squares) basis  coefficients using a dense B-spline basis with equally-spaced knots ({\bf bottom  left}); and the  second differences of the basis coefficients ({\bf bottom right}).}
    \label{illustrations_example}
\end{figure}

Our strategy marries a particular choice of basis functions with a locally adaptive shrinkage prior. Specifically, we select a B-spline basis with a moderate number of equally-spaced knots. B-splines are numerically stable with convenient computational properties, in part due to the local compactness and ordering among the basis functions. These properties further motivate and enable our prior specification for the basis coefficients. As an illustrative example, consider the B-spline basis coefficients for the nonlinear function in Figure~\ref{illustrations_example}. The function is smooth yet features two regions with rapid changes, which are highlighted by the plot of $\beta''$. The B-spline basis coefficients $\bm B^*$ (determined via least squares for this illustration) offer several suggestions for an ideal prior. First, the coefficients inherit an ordering similar to that in the original domain $\mathcal{T}$. Thus, neighboring coefficients should be shrunk together to encourage smoothness. Next, the second differences of the coefficients, $\Delta^2 B_k^* = \Delta B_k^* - \Delta B_{k-1}^*$ with $\Delta B_k^* = B_{k+1}^* - B_{k}^*$, closely resemble the second derivatives $\beta''$. This observation has motivated P-splines \citep{marx1999generalized}, which imitate the familiar roughness penalty $\int_\mathcal{T}\{ \beta''(t)\}^2\ dt$ with the coefficient analog $\sum_k (\Delta^2 B_k^*)^2$. However, this \emph{global} penalty ignores the final critical observation in Figure~\ref{illustrations_example}: the smooth periods of $\beta$ correspond to zeros in $\Delta^2 B_k^*$, while the rapidly-changing periods exhibit \emph{volatility clustering}, i.e., consecutive sequences of large absolute values. More specifically, the shrinkage in the peaked regions (around the 50-80th and the 120-150th coefficients) should not be as aggressive as in the flat regions. Classical  P-splines cannot capture this behavior: the rate of shrinkage is global across all $k$ and $\mathcal{T}$. Thus, adequate B-spline modeling of functions with both smooth and rapidly-changing features requires a prior that (i) encourages smoothness via near-sparsity of  $\Delta^2 B_k^*$ and (ii) admits local adaptivity via dynamic volatility modeling.

%Note that in practice, the regression function $\beta$ is not directly observable as in Figure~\ref{illustrations_example}. 

%%%%%%%%%%%

Motivated by these considerations, we propose the following locally adaptive shrinkage prior on the regression coefficient function:
\begin{align}
\beta(t) &= \sum_{k=1}^{K_B}B_k^* \psi_k(t) 
 \label{b-spline sofr}\\
\Delta ^2 B_{k}^*\mid \lambda_k &\stackrel{\text{indep}}{\sim} \mathcal{N}(0,\lambda_k^2), \quad \{\lambda_k\} \sim \text{DHS} \label{dhs_Bspline}
\end{align}
where $\{\psi_k\}_{k=1}^{K_B}$ is a collection of equally-spaced B-splines and DHS refers to the dynamic horseshoe prior \citep{kowal2019dynamic}. The local scale parameters $\lambda_k$ determine the smoothness of the function over the (compact) support of $\psi_k$: when $\lambda_k$ is small, the function $\beta$ is approximately locally linear; when $\lambda_k$ is large, the function $\beta$ admits large changes in the slope. This local adaptivity is enabled by the basis-specific scales $\lambda_k$. By comparison, classical P-splines apply global smoothness via a common scale $\lambda_k = \lambda$. 

\begin{figure}
    \centering
     \includegraphics[width = \textwidth]{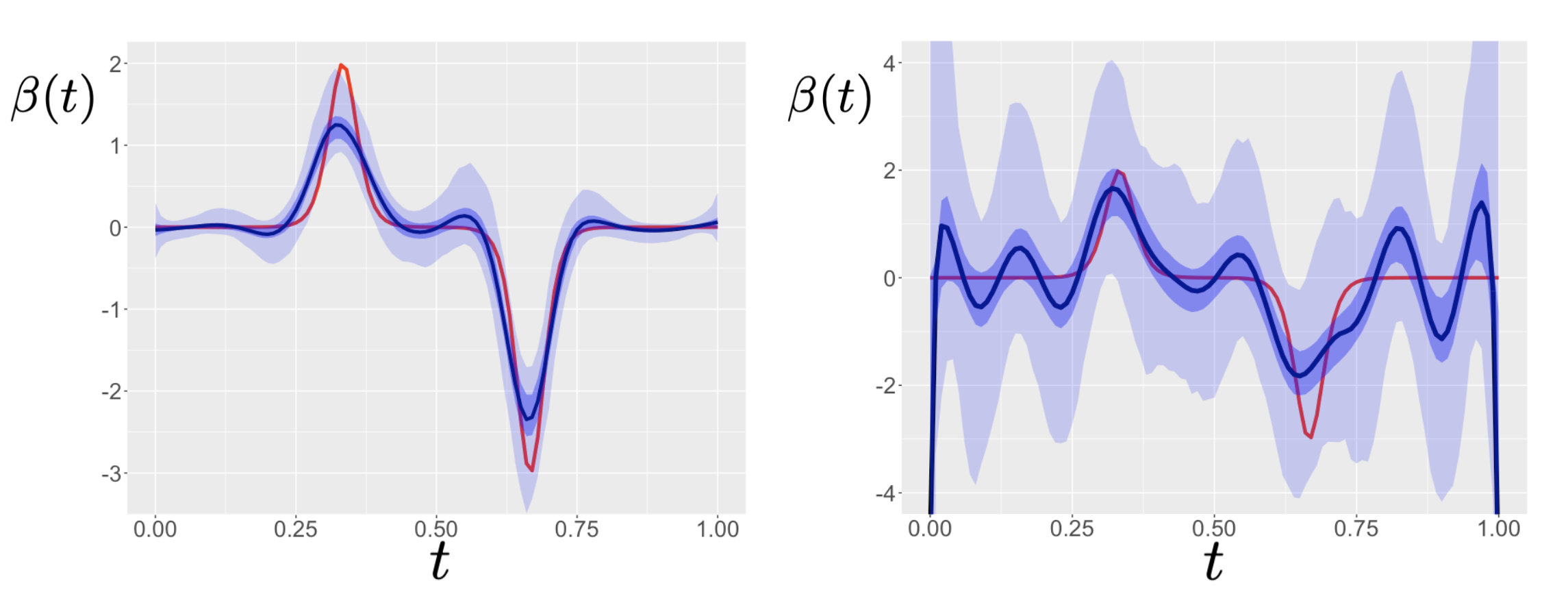}
    \caption{ True curve (red) with the posterior mean (dark line) and 50\%  (dark shade) and  95\% (light shade) credible intervals for the proposed method  ({\bf left}) and the local P-spline competitor ({\bf right}). Both the \emph{dynamic} and \emph{shrinkage} properties of the proposed prior are necessary for accurate point estimation and precise uncertainty quantification.}
    \label{quick_example}
\end{figure}

Relative to the horseshoe prior \citep{carvalho2010horseshoe}, the dynamic horseshoe prior offers key advantages for adaptive function estimation. The horseshoe prior assumes \emph{independent} half-Cauchy priors for $\lambda_k$, which does not account for the volatility clustering observed in Figure~\ref{illustrations_example}. Informally, non-dynamic shrinkage priors do not incorporate information regarding the shrinkage behavior of neighboring regions, which produces inferior estimates and uncertainty quantification  (see Figure~\ref{quick_example}). The dynamic horseshoe prior resolves these issues using a volatility model: 
\begin{eqnarray}
h_k \coloneqq \log (\lambda_k^2), \quad h_{k+1} = \mu_h + \phi(h_k-\mu_h) + \eta_{k+1}, \quad  \eta_k \stackrel{iid}{\sim} Z(1/2, 1/2, 0, 1) \label{DHS}
\end{eqnarray}
where $Z (a, b, 0,1)$ denotes the $Z$-distribution with density 
$ [z] = \{ B(a,b)\}^{-1}\exp\{z\}^a [1 + \exp\{z\}]^{-(a + b)}$  
and $B(\cdot,\cdot)$ is the beta function. The dynamic horseshoe prior models the log-variances of the second-differenced basis coefficients with an autoregressive model of order one, which resembles classical Bayesian volatility models for time series analysis \citep{kim1998stochastic}. The key distinctions here are (i) the presence of the $Z$-distribution and (ii) the role of the coefficient indices $k$. First, \cite{kowal2019dynamic} showed that many common shrinkage priors expressed via $\lambda_k$ can be represented on the log-scale with a $Z$-distribution; see Table~\ref{ibDist}. When there is no autoregressive behavior $\phi = 0$, the prior \eqref{DHS} with $a = b = 1/2$ exactly reproduces the horseshoe prior. As such the dynamic horseshoe is capable of providing both aggressive shrinkage and persistence of large signals, which here corresponds to local smoothness and rapidly-changing features in $\beta$, respectively. 

\begin{table}
\centering 
\caption{Each of these priors is reproduced by a $Z$-distribution on the log-variance.  \label{ibDist}}
\begin{tabular}{ll}
\hline
$a = b = 1/2 $ &Horseshoe Prior \citep{carvalho2010horseshoe} \\ 
$a = 1/2 , b=1$ & Strawderman-Berger Prior   \citep{strawderman1971proper,berger1980robust} \\ 
$a= 1, b = c-2, c > 0 $ &Normal-Exponential-Gamma Prior  \citep{griffin2005alternative} \\ 
$a = b  \rightarrow 0  $ & (Improper) Normal-Jeffreys' Prior  \citep{figueiredo2003adaptive} \\ 
\hline
\end{tabular}
\end{table}

Second, the "time" index for the volatility model is $k$, which corresponds to the second differenced basis coefficient $\Delta^2 B_k^*$. This modeling structure is appropriate due to the local compactness and ordering of the (equally-spaced) B-spline basis functions (Figure~\ref{illustrations_example}).

To demonstrate the importance of both the \emph{dynamic} and \emph{shrinkage} aspects of the prior for SOFR, we consider a brief example with simulated data. Data from the SOFR model \eqref{SOFR} are generated for $n=500$ observations with a moderate signal-to-noise ratio (SNR = 5) using the nonlinear function from Figure~\ref{illustrations_example} for the true regression coefficient function; additional details are provided in Section~\ref{sims}. To compare with the proposed approach, we consider a variation of \eqref{b-spline sofr}--\eqref{dhs_Bspline} that instead uses independent and diffuse inverse-Gamma priors on $\{\lambda_k^2\}$. This \emph{local P-spline} competitor includes local scale parameters, but fails to provide either the aggressive shrinkage or the dynamics of the proposed approach.
The posterior means and 50\% and 95\%  pointwise credible intervals for $\beta$ are presented in Figure~\ref{quick_example}. Clearly, this example is highly challenging: the nonlinear regression function includes both flat and rapidly-changing features, and is not directly observable and must be inferred via the regression model \eqref{SOFR}. Most striking, the dynamic shrinkage provides better point estimation, especially in the smooth regions, and significantly more narrow interval estimates. By comparison, the local P-spline incorrectly estimates oscillations that are not present in the true function and produces credible intervals that are far too wide to be useful in practice.

The proposed \emph{Bayesian adaptive scalar-on-function regression} (BASOFR) model   \eqref{SOFR} and \eqref{mlr}--\eqref{DHS} is completed  by specifying priors  on the remaining parameters  in \eqref{SOFR} and  \eqref{DHS}. 
By default, we assume the diffuse priors $[\mu]\propto 1$ and $[\sigma^{-2}] \sim \mbox{Gamma}(0.01, 0.01)$. For the boundary coefficients in \eqref{dhs_Bspline}, we include the prior $[B_1^*, B_{K_B}^* \mid \lambda_0]\sim \mathcal{N}(0, \lambda_0^2)$ indepenently with $\lambda_0^{-2} \sim \mbox{Gamma}(0.01, 0.01)$, which guards against excessively wide interval estimates of $\beta$ near the boundaries of   $\mathcal{T}$.  For the dynamic horseshoe parameters, we follow  \cite{kowal2019dynamic} and assume  $[\exp(\mu_h/2)]  \sim C^+(0,1)$, which  corresponds to the global scale parameter in the non-dynamic horseshoe ($\phi  = 0$) and $[(\phi + 1)/2] \sim \mbox{Beta}(10, 2)$, which encourages persistence in the log-volatility but maintains stationarity via $\vert \phi \vert  < 1$. 

Posterior inference under this model is available using an efficient Gibbs sampler that cycles through the basis coefficients $\{B_k^*\}$ in (\ref{b-spline sofr}),   the log-volatilities $\{h_k\}$ in (\ref{DHS}), the autoregressive parameters $ \{\mu_h,\phi\}$, and the variance component $\sigma^2$ in \eqref{SOFR}. The crucial features of our sampling algorithm are (i) the full conditional distribution of $\{B_k^*\}$ is $K_B$-dimensional Gaussian, which can be sampled efficiently and used to update $\beta(t) = \sum_{k=1}^{K_B}B_k^* \psi_k(t)$ for any $t \in \mathcal{T}$, and (ii) the log-volatilities $\{h_k\}$ can be sampled using a fast $\mathcal{O}(K_B)$ algorithm that relies on Gaussian parameter expansions and banded precision matrices; the details are provided in the supplement. 

\begin{figure}
    \centering
    \includegraphics[width=0.49\textwidth]{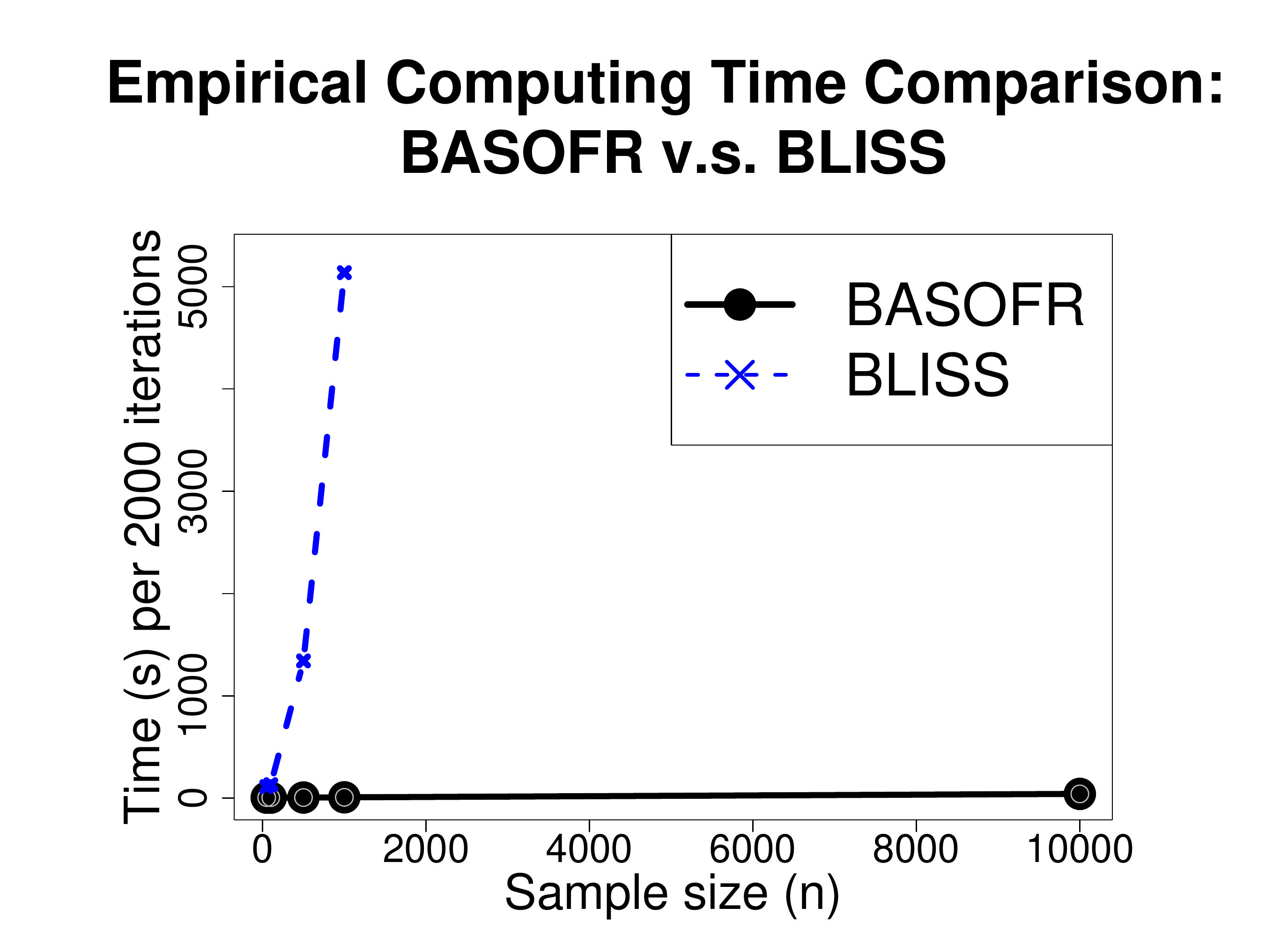}
    \includegraphics[width=0.49\textwidth]{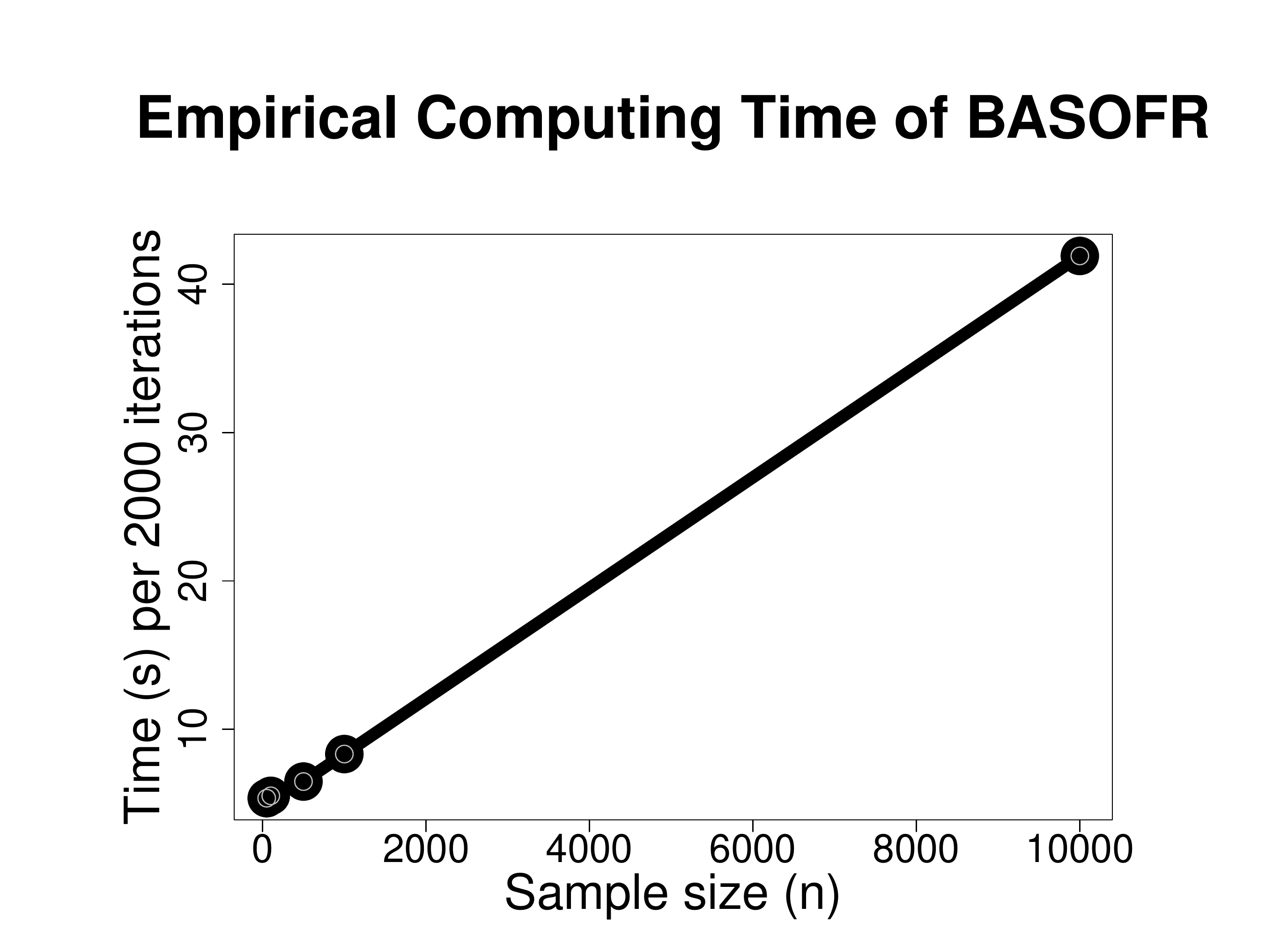}
    \caption{The empirical computational time for BASOFR and BLISS (\textbf{left}) and BASOFR (\textbf{right}) for $n \in\{50,100,500,1000,10000\}$ in seconds per 2000 iterations (using R on a Windows PC, 2.79 GHz AMD EPYC-Rome). %The computation time is averaged in 10, 5, 5, 2, and 1 replicates for $n=50,100,500,1000,10000$, respectively.  
     }
    \label{fig:comptime_vs}
\end{figure}

To highlight the computational scalability, we compare the empirical computing time for BASOFR against  BLISS \citep{grollemund2019bayesian} in Figure~\ref{fig:comptime_vs}. We vary the sample sizes $n$ and use the true regression function from Figure~\ref{quick_example}. To ensure favorable conditions for BLISS, we fix the number of local levels at two so that the computation time for BIC model selection is not included; thus, these computing times \emph{underestimate} the usual computational burden of BLISS. Nonetheless, it is clear that BLISS does not scale to even moderate sample sizes $(n > 1000)$, while the proposed algorithm scales approximately linearly in $n$. Since our application features $n \approx 100,000$, such scalability is essential. 

\section{Decision analysis for window selection in SOFR}\label{PS}
The BASOFR proposed  in \eqref{SOFR} and \eqref{mlr}--\eqref{DHS} offers several key benefits for Bayesian SOFR, including more adaptive  point estimation, more precise uncertainty quantification, and scalable computing capabilities (see Section~\ref{sims}). Despite these advantages, the posterior distribution of $\beta$ alone cannot select critical windows of susceptibility:  selection must be considered carefully and requires a decision analysis. Further, the local adaptivity induced by the prior does  \emph{not} resolve the interpretability issues noted in \eqref{n-interp-1}--\eqref{n-interp-2}. This challenge permeates Bayesian SOFR and Bayesian inference more broadly: models may produce accurate data-generating processes, yet the parameters of those  models do not necessarily offer the most convenient interpretations.

To address these challenges, we introduce \emph{posterior summarization} tools for SOFR. Informally, the strategy adopts a two-stage approach: first, we estimate an adequate Bayesian SOFR model, such  as BASOFR; second, we  extract more interpretable point summaries from the model. These summaries are designed to replace more common point estimates, such as the posterior mean of $\beta(t)$, but customized to provide window selection and to mitigate the challenges observed in \eqref{n-interp-1}--\eqref{n-interp-2}. Specifically, we target  \emph{locally constant} (or stepwise) point estimates of the form $\hat \beta(t) = \hat \delta_k$  for $t \in \mathcal{T}_k$  and $\{\mathcal{T}_k\}_{k=1}^K$ a partition of $\mathcal{T}$. Importantly, the key SOFR term  simplifies: 
\begin{equation}
    \label{local}
    \int_{\mathcal{T}} X_i( t)\hat  \beta (t)\ dt = \sum_{k=1}^K X_i(\mathcal{T}_k) \hat\delta_k
\end{equation}
where $X(\mathcal{T}_k)\coloneqq \int_{\mathcal{T}_k} X(t) \ dt$ is the aggregated trajectory over $\mathcal{T}_k$. The partition $\{\mathcal{T}_k\}_{k=1}^K$  provides  window selection, i.e., the regions of $\mathcal{T}$ that  are most important for predicting $y$. Hence, it directly targets our goal of identifying the susceptible windows of $\mbox{PM}_{2.5}$ exposure during gestation, specifically linked to educational outcomes.  These coefficients are also arguably  more interpretable than $\beta(t)$: $\hat \delta_k$ estimates the change in the expectation of $y$ for a one-unit increase in the \emph{aggregated} trajectory $X(\mathcal{T}_k)$ while  holding the remaining \emph{aggregated} trajectories $\{X(\mathcal{T}_j)\}_{j  \ne k}$ constant.
As  an added benefit, this representation significantly reduces storage requirements: point predictions can be computed using the aggregated trajectories $\{X(\mathcal{T}_k)\}_{k=1}^K$ instead of the entire trajectories $\{X(t)\}_{t \in \mathcal{T}}$. 

%In the context of  our motivating data example, suppose $\mathcal{T}_1$ corresponds to the third trimester during gestation; interpreting $\delta_1$ only requires considering a one-unit increase in $\mbox{PM}_{2.5}$ exposure during the third trimester while holding constant the exposures in the remaining partition segments. By comparison, the analogous interpretation of $\beta(t^*)$ requires considering a one-unit increase in $\mbox{PM}_{2.5}$ exposure \emph{only} on day $t^*$---and no others. More broadly, the aggregated  trajectories  $\{X_i(\mathcal{T}_k)\}_{k=1}^K$ typically demonstrate substantially lower correlations than the original trajectories  $\{X_i(t)\}_{t \in \mathcal{T}}$. Hence, the "all else equal" interpretations strongly benefit from the reduced correlations. 

To extract these summaries, we adopt a decision analysis approach and use the \emph{acceptable families} of \cite{kowal2021fast} to compare partitions. Specifically, consider the following predictive loss function for $\delta$, omitting the scalar covariates $\bm z_i$ for now:
\begin{equation}
        \tilde L_\lambda(\delta) = n^{-1} \sum_{i=1}^{ n} \Vert (\tilde y_i-\mu) - \sum_{k=1}^K X_i(\mathcal{T}_k) \delta_k \Vert_2^2 + \lambda \sum_{k=2}^K \vert \delta_k-  \delta_{k-1}\vert, \label{loss_locally_constant}
\end{equation}
where each $\tilde y_i$ is a posterior predictive variable at $X_i$ under the BASOFR model. The loss function $\tilde L_\lambda(\delta)$ combines a "goodness-of-fit" component with an $\ell_1$-penalty on the increments $ \delta_k - \delta_{k-1} $ to encourage fewer change points in $\delta_k$. The loss function may be further augmented with an $\ell_1$-penalty on $\delta_k$ or other thresholding to encourage additional sparsity. 

Since this loss inherits a posterior (predictive) distribution via $\{\tilde y_i\}_{i=1}^n$ and $\mu$, Bayesian decision analysis proceeds by integrating over the posterior (predictive) distribution and minimizing the resulting quantity: $\hat \delta_\lambda \coloneqq \arg \min_{\delta} \mathbb{E}_{[ \tilde y , \mu\mid y]} \tilde L_\lambda(\delta)$, which simplifies to
\begin{equation}
%\label{post_expectation_loss}
    \hat \delta_\lambda %& \coloneqq \arg \min_{\delta} \mathbb{E}_{[ \tilde y , \mu\mid y]} \tilde L_\lambda(\delta) \\
    \label{fusedlasso} 
    = \arg \min_{\delta} \Big\{ n^{-1} \sum_{i=1}^{ n}\Vert (\hat y_i - \hat \mu) - \sum_{k=1}^K X_i(\mathcal{T}_k)\delta_k \Vert_2^2 + \lambda \sum_{k=2}^K \vert \delta_k - \delta_{k-1}\vert \Big\}
\end{equation}
(assuming $\mathbb{E}_{[ \tilde y , \mu\mid y]} \Vert \tilde y_i - \mu \Vert^2 < \infty$; \citealp{kowal2021fast}), 
where $\hat y_i \coloneqq \mathbb{E}_{[\tilde y_i \mid y]} \tilde y_i$ and $\hat \mu \coloneqq \mathbb{E}_{[\mu \mid y]} \mu$ are posterior predictive expectations under the BASOFR model. Crucially, the \emph{optimal Bayes action} $\hat \delta_\lambda$ is a "fit-to-the-fit" using pseudo-data $\hat y_i - \hat \mu =  \int_{\mathcal{T}} X_i( t)\hat \beta (t)\ dt$ for $\hat \beta(t) = \mathbb{E}_{[\beta \mid y]} \beta(t)$ and covariates $\{X_i(\mathcal{T}_k)\}_{k=1}^K$. As such, $\hat \delta_\lambda$ seeks to simplify point estimation not by targeting $\hat \beta(t)$ directly, but rather optimizing for the point predictions generated by $\hat \beta(t)$ under the SOFR model. Given these point predictions, the solution in \eqref{fusedlasso} is readily computed using existing software, such as the \texttt{R} package \texttt{genlasso}    \citep{tibshirani2011solution}. 

The decision-analytic optimality of $\hat \delta_\lambda$   is valid only for a fixed $\lambda$, which controls the number of partitions (or steps) in the locally constant estimator $\{\hat \delta_k\}$. Hence, further comparisons are required across the path of $\lambda$ values. We consider two metrics for each $\hat\delta$: the \emph{empirical} mean squared error 
\begin{equation}
    \label{empirical}
    {\mathcal{E}}_{\lambda} = \frac{1}{n} \sum_{i=1}^{n} \big\{ (y_i - \hat \mu) - \int_{\mathcal{T}_i}  X_i(t)\hat \delta_{\lambda}(t)\ dt \big\}^2
\end{equation}
and the \emph{predictive} mean squared error $\widetilde{\mathcal{E}}_{\lambda} \coloneqq \tilde L_0(\hat \delta_\lambda)$ via \eqref{loss_locally_constant}, which replaces $y_i$ with $\tilde y_i$ and $\hat \mu$ with $\mu$ in \eqref{empirical}.  
%\begin{equation}
  %  \label{predictive}
%    \widetilde{\mathcal{E}}_{\lambda} = \frac{1}{n} \sum_{i=1}^{n} \big\{ (\tilde y_i - \mu) - \int_\mathcal{T}  X_i(t)\hat \delta_{\lambda}(t)\ dt \big\}^2.
%\end{equation}
Both metrics are important:    ${\mathcal{E}}_{\lambda} $ provides an empirical point summary of the predictive accuracy, while $\widetilde{\mathcal{E}}_{\lambda}$ inherits a posterior predictive distribution under the BASOFR model via $\tilde y$ and $\mu$. 

The uncertainty quantification provided by the predictive version $\widetilde{\mathcal{E}}_{\lambda}$ is valuable for comparing across approximations of varying complexities $\lambda$. In particular, a primary drawback of the locally constant representation \eqref{local} is the potential for instability, i.e., distinct partitions $\{\mathcal{T}_k\}$ and  $\{\mathcal{T}_k'\}$ that produce similar predictive performance yet differ in their identification of the important windows of $\mathcal{T}$. This issue is not unique to our posterior summarization strategy, but persists more broadly for estimators of the form \eqref{local}. To address this instability, we leverage the uncertainty quantification from $\widetilde{\mathcal{E}}_{\lambda}$ to construct the \emph{acceptable family} \citep{kowal2021fast}, which collects the approximations $\hat \delta_\lambda$ that offer "near-optimal" predictive performance: 
\begin{equation}\label{accept_family}
    \mathcal{A}_\varepsilon  \coloneqq \big\{\lambda: \mathbb{P}_{\mathcal{M}}(\widetilde {\mathcal{D}}_{\lambda} < 0)\geq \varepsilon\big\}, \quad  \varepsilon \in [0,1]
\end{equation}
where $\widetilde{\mathcal{D}}_{\lambda}\coloneqq 100 \times (\widetilde {\mathcal{E}}_{\lambda}-\widetilde {\mathcal{E}}_{\lambda_{\min}})/\widetilde {\mathcal{E}}_{\lambda_{\min}}$ is the percent increase in predictive mean squared error relative to the empirical loss minimizer $\lambda_{\min} \coloneqq \arg\min_\lambda \mathcal{E}_\lambda$. Informally, $ \mathcal{A}_\varepsilon$ collects all approximations $\hat \delta_\lambda$ for which the predictive performance matches or exceeds that of $\hat\delta_{\lambda_{\min}}$ with at least $\varepsilon$ probability under the BASOFR model $\mathcal{M}$. Equivalently, $\lambda \in   \mathcal{A}_\varepsilon$ if and only if a lower $(1 - \varepsilon)$ posterior prediction interval for $\widetilde{\mathcal{D}}_{\lambda}$ includes zero \citep{kowal2021fast}. The  acceptable family has been applied for targeted prediction \citep{kowal2021fast}, variable  selection \citep{kowal2021bayesian}, subset selection \citep{kowal2022bayesian}, and selection in mixed effects models \citep{kowal2021subset}. By default, we select $\varepsilon = 0.10$; smaller values expand the acceptable family, but results are generally robust to moderate changes in $\varepsilon$ \citep{kowal2021fast,kowal2021bayesian,kowal2022bayesian}. We focus on the \textit{simplest} member of the acceptable family, i.e. the locally constant point estimate with the fewest changes in the local level (yet still satisfies \eqref{accept_family}).

To incorporate the scalar covariates $\bm z_i$ in \eqref{SOFR}, we replace $(\tilde y_i - \mu)$ with $(\tilde y_i - \mu - \bm z_i'\bm\alpha)$ in \eqref{loss_locally_constant} and $\widetilde{\mathcal{E}}_{\lambda}$  and $(\hat y_i - \hat \mu)$ with $(\hat y_i - \hat \mu - \bm z_i'\bm{\hat\alpha})$ for $\bm{\hat \alpha} \coloneqq \mathbb{E}_{[\alpha \mid y]} \bm \alpha$ in \eqref{fusedlasso} and \eqref{empirical}. The resulting optimal point estimates  $\hat \delta_\lambda$ now  account for the scalar covariates $\bm z_i$, while the posterior predictive quantities  \eqref{loss_locally_constant} and $\widetilde{\mathcal{E}}_{\lambda}$ include the uncertainty due to the model parameters $\bm \alpha$. 

\section{Simulation study}\label{sims} 

We conduct two simulation studies: one that evaluates the BASOFR model against other SOFR models (Section~\ref{sims_1}) and one that assesses the decision analysis approach for window selection (Section~\ref{sims_2}), both using simulated datasets that resemble the NC data in our application study. Since the daily $\mbox{PM}_{2.5}$ trajectories are seasonal (see Section~\ref{apps}), we generate functional covariates $\{X_i\}_{i=1}^n$ with a seasonal pattern: each $X_i$ follows a Gaussian process with mean function $\mu_i(t) =  \sin (2\pi t/T  + \phi_i)$ and covariance function $\text{Cov}(X_i(t), X_i(t'))= \sigma_x^2 \exp\{-(t-t')^2/(2\ell^2)\}$. The period parameter $T$ is fixed to induce an annual pattern 
($T = 365/ \text{maximal gestational length in days} = 365/295$), the offset $\phi_i \sim \mbox{Unif}(0,1)$ represents births at different times of year, and  $\sigma_x = 0.7$ and $\ell = 0.01$ are chosen to visually resemble the $\mbox{PM}_{2.5}$ exposure curves. Each functional covariate is evaluated on  a common and regular grid $(0,0.01,0.02,\ldots,1)$. The supplementary material includes results for smooth yet non-seasonal functional covariates, which is an easier setting for estimation and inference yet produces the same comparative results as those below. 

\subsection{BASOFR for point estimation and uncertainty quantification}\label{sims_1} 
For a challenging estimation and inference scenario, we adopt the regression coefficient function in Figure \ref{illustrations_example}, which presents both smooth and rapidly-changing features. The simulated datasets vary in the sample sizes and signal-to-noise ratios (SNR), with the SNR decreasing as $n$ increases:  $(n, \mbox{SNR}) \in \{(50, 10), (100, 7), (500, 5), (10,000, 0.5)\}$. Using the aforementioned seasonal functional covariates, the  response variables are simulated from \eqref{SOFR} with $\mu=0$ and  $\sigma$ determined based on the SNR, and the process is repeated to generate 50 datasets. 
%For each design, we generate 50 simulated datasets. 

To compete with the BASOFR, we include BLISS  \citep{grollemund2019bayesian} and two Bayesian variations of the B-spline model \eqref{b-spline sofr}--\eqref{dhs_Bspline}. BLISS estimation follows the default recommendations to  fit separate models with 1 to 5 levels and  select the model with the lowest BIC. Due to the high computational cost  (see Figure \ref{fig:comptime_vs}), we only include BLISS  for $n\in \{50,100\}$. Next, we modify \eqref{b-spline sofr}--\eqref{dhs_Bspline} to include a global smoothness parameter  
$\lambda_k = \lambda$ and a diffuse inverse-Gamma  prior on $\lambda^2$ (P-spline) 
and the local P-spline model from Figure~\ref{quick_example} with  independent and diffuse inverse-Gamma priors for each $\lambda_k^2$. Each model provides a  point estimate of $\beta$ via the posterior expectation and uncertainty quantification for $\beta$ via 95\% pointwise credible intervals. 

\begin{figure}
    \centering
    \includegraphics[width=0.49\textwidth]{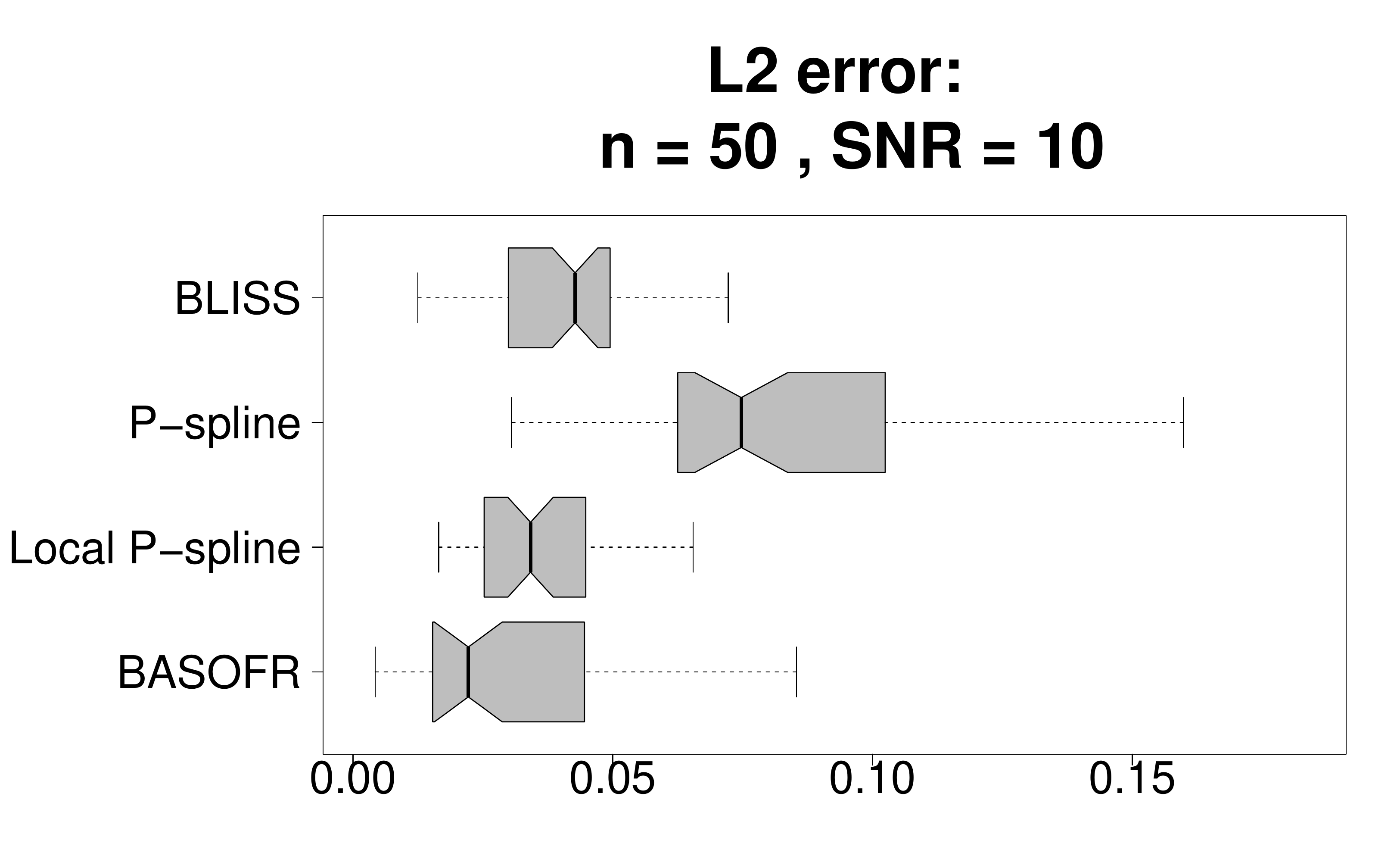}
    \includegraphics[width=0.49\textwidth]{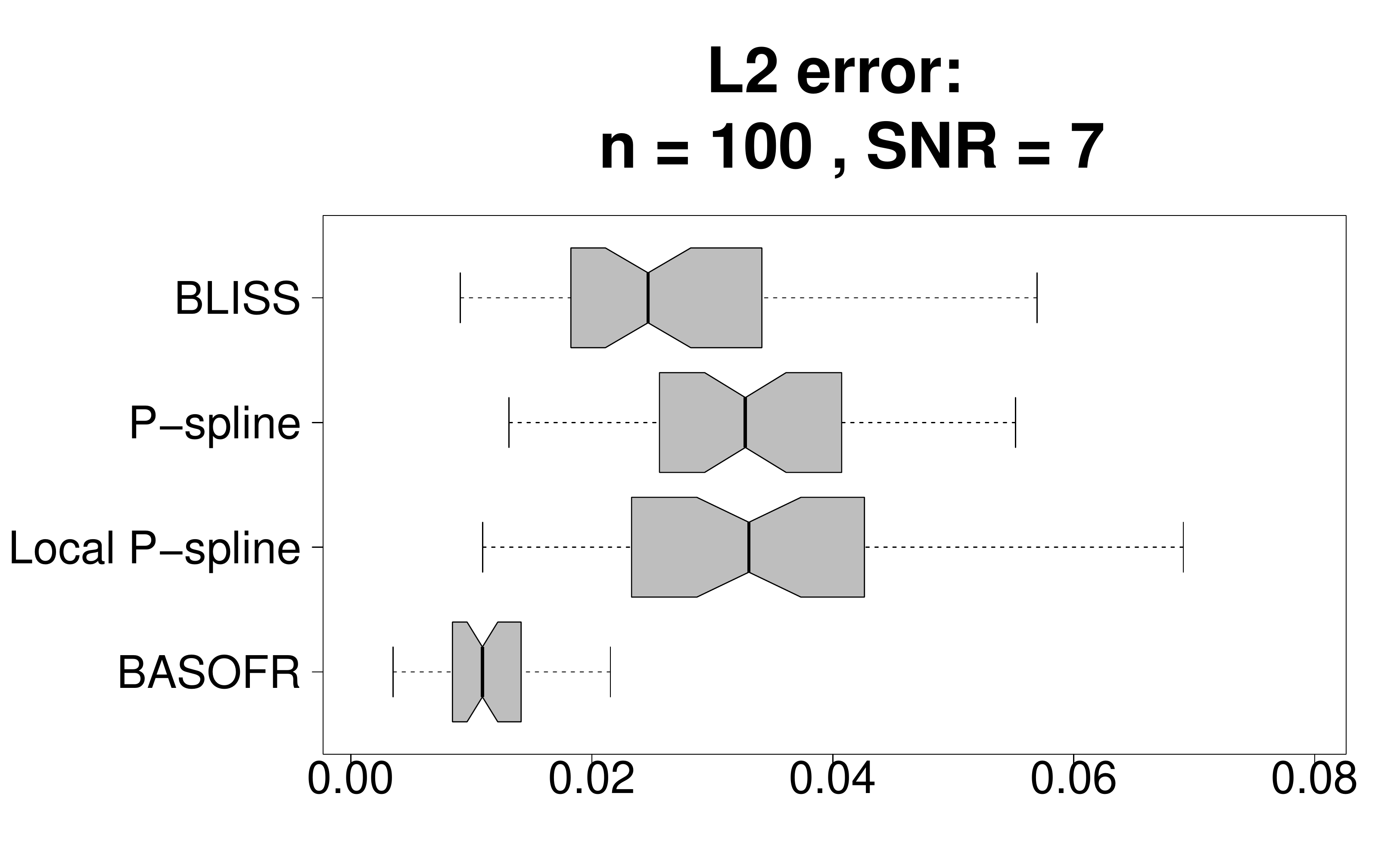} 
    \includegraphics[width=0.49\textwidth]{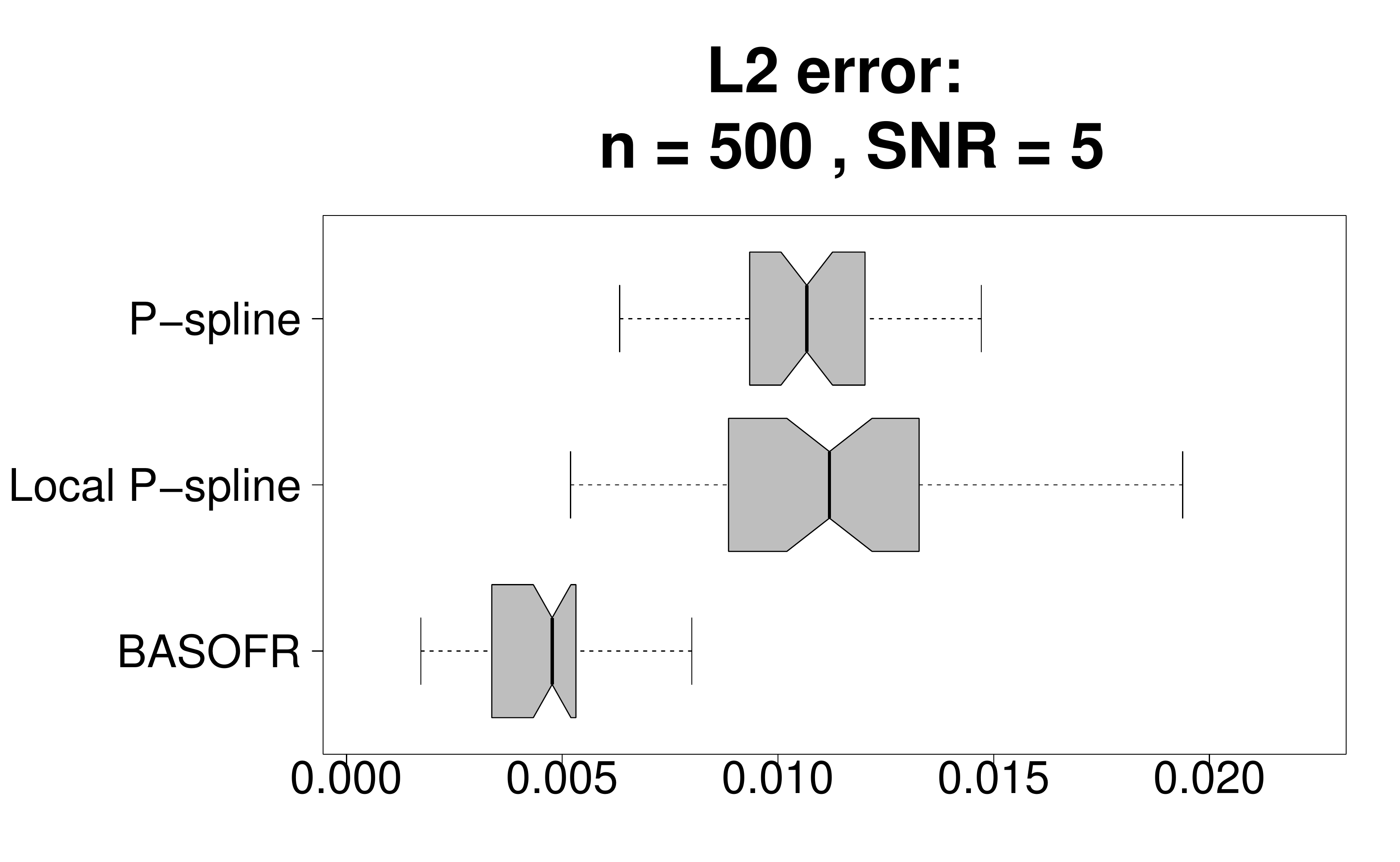}
    \includegraphics[width=0.49\textwidth]{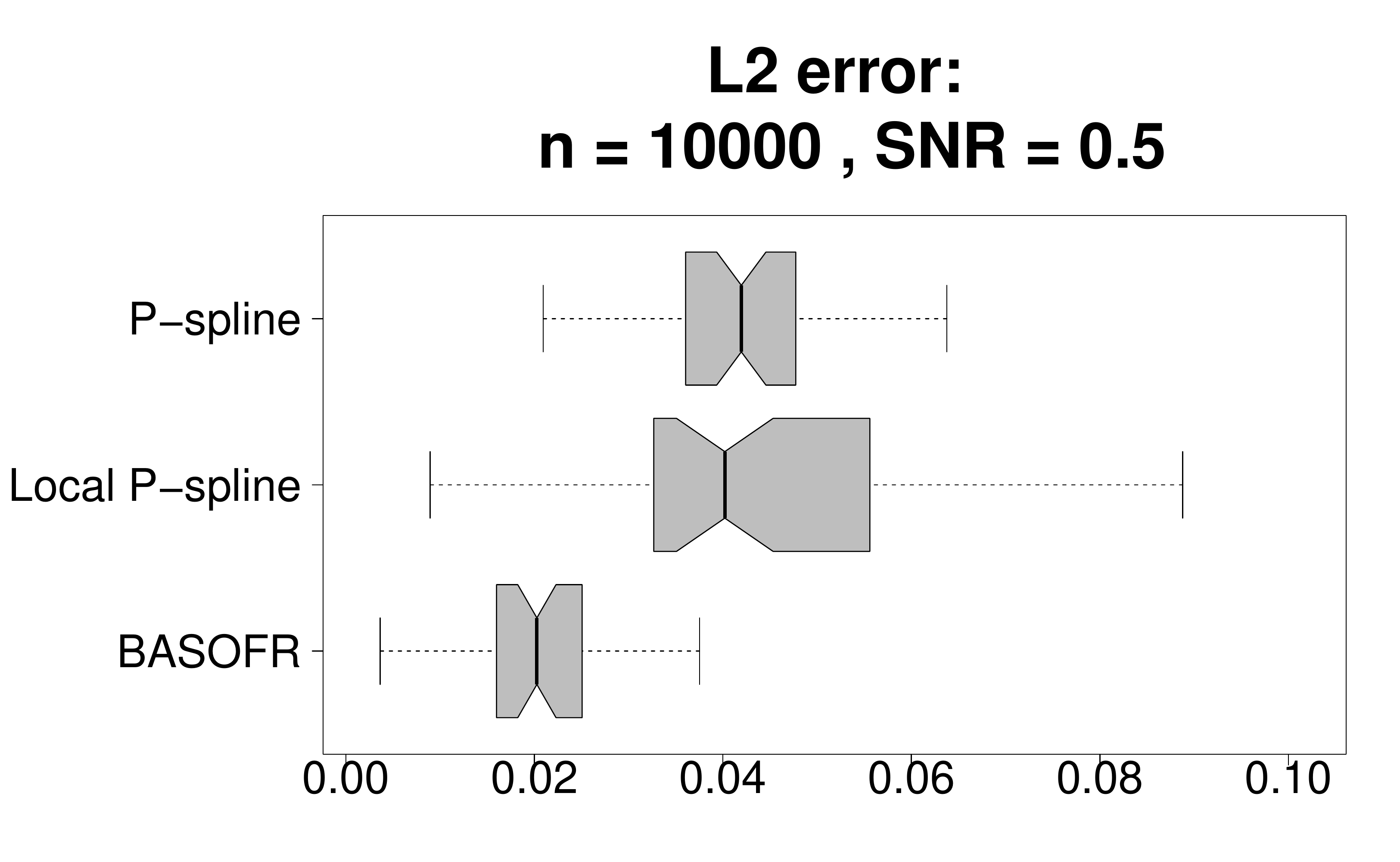}
    \caption{$L_2$-error for  estimating the true regression function.  BASOFR offers the most accurate point estimation with larger gains over competing methods as the sample size increases.} %The $(n, \mbox{SNR}) = (500, 5)$ case is nearly identical to $(n, \mbox{SNR}) = (10000, 0.5)$ and is  omitted.}
    \label{fig:l2error_seasonal}
\end{figure}

\begin{figure}
    \centering
     \includegraphics[width=0.49\textwidth]{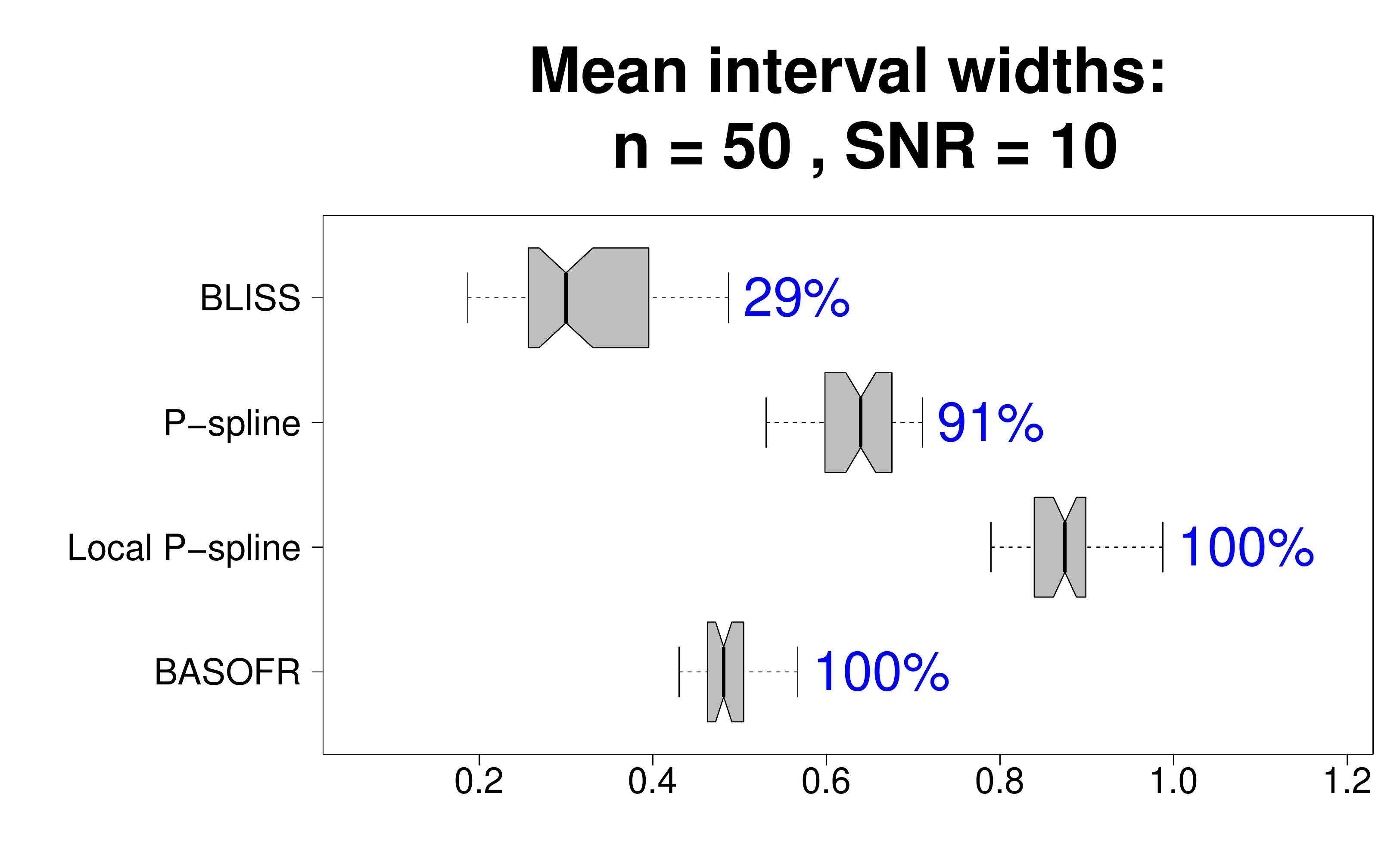}
    \includegraphics[width=0.49\textwidth]{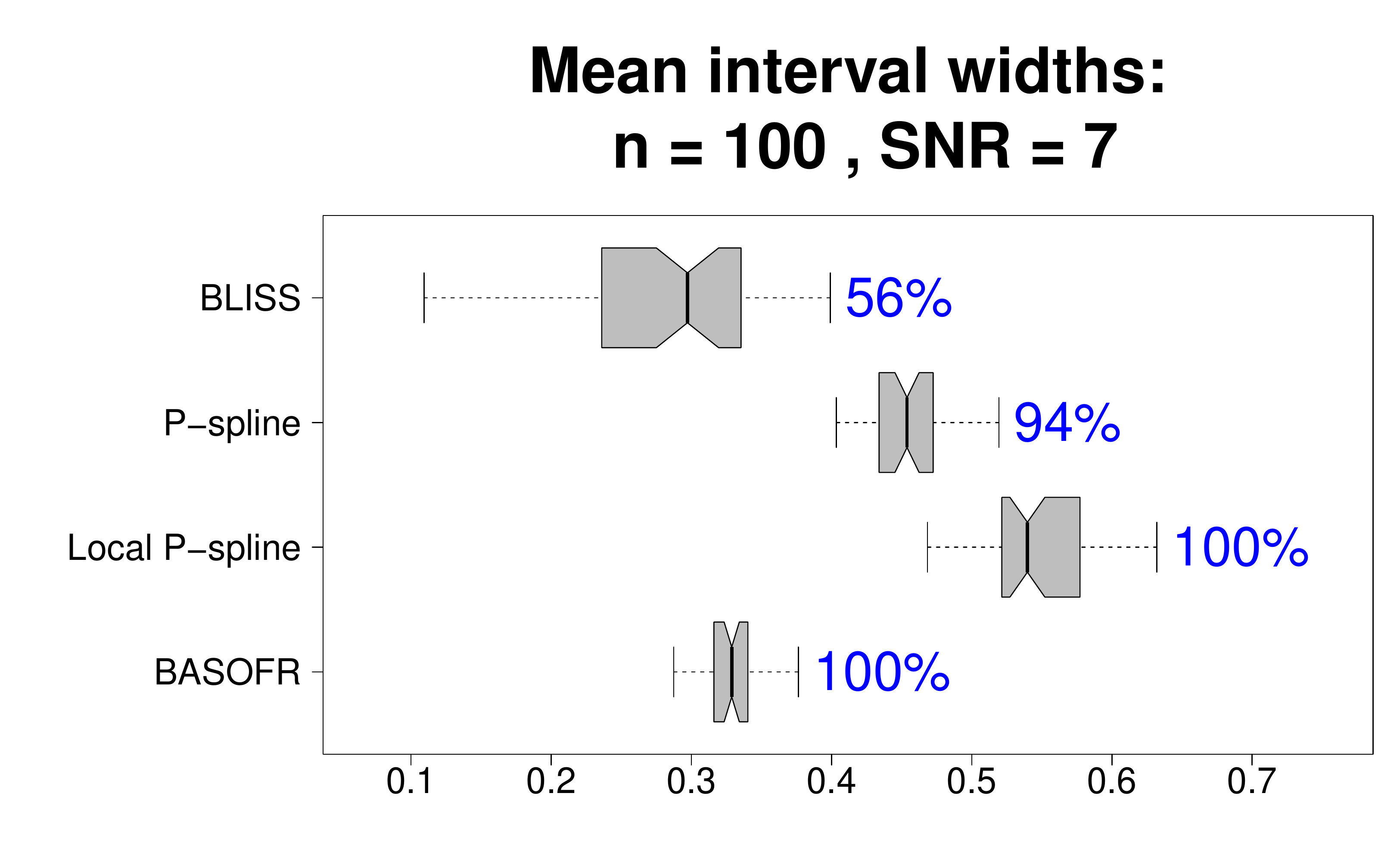}
    \includegraphics[width=0.49\textwidth]{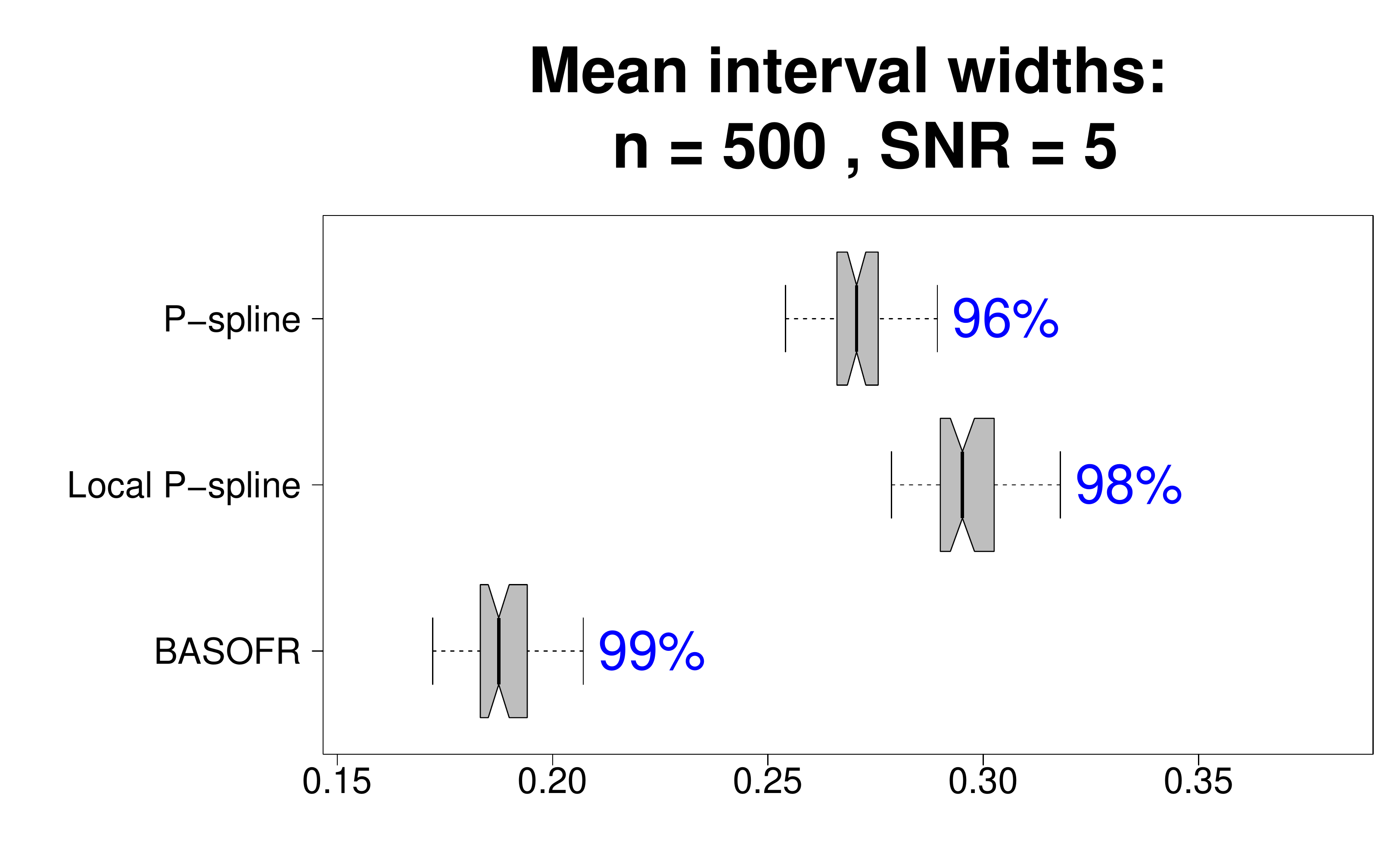}
    \includegraphics[width=0.49\textwidth]{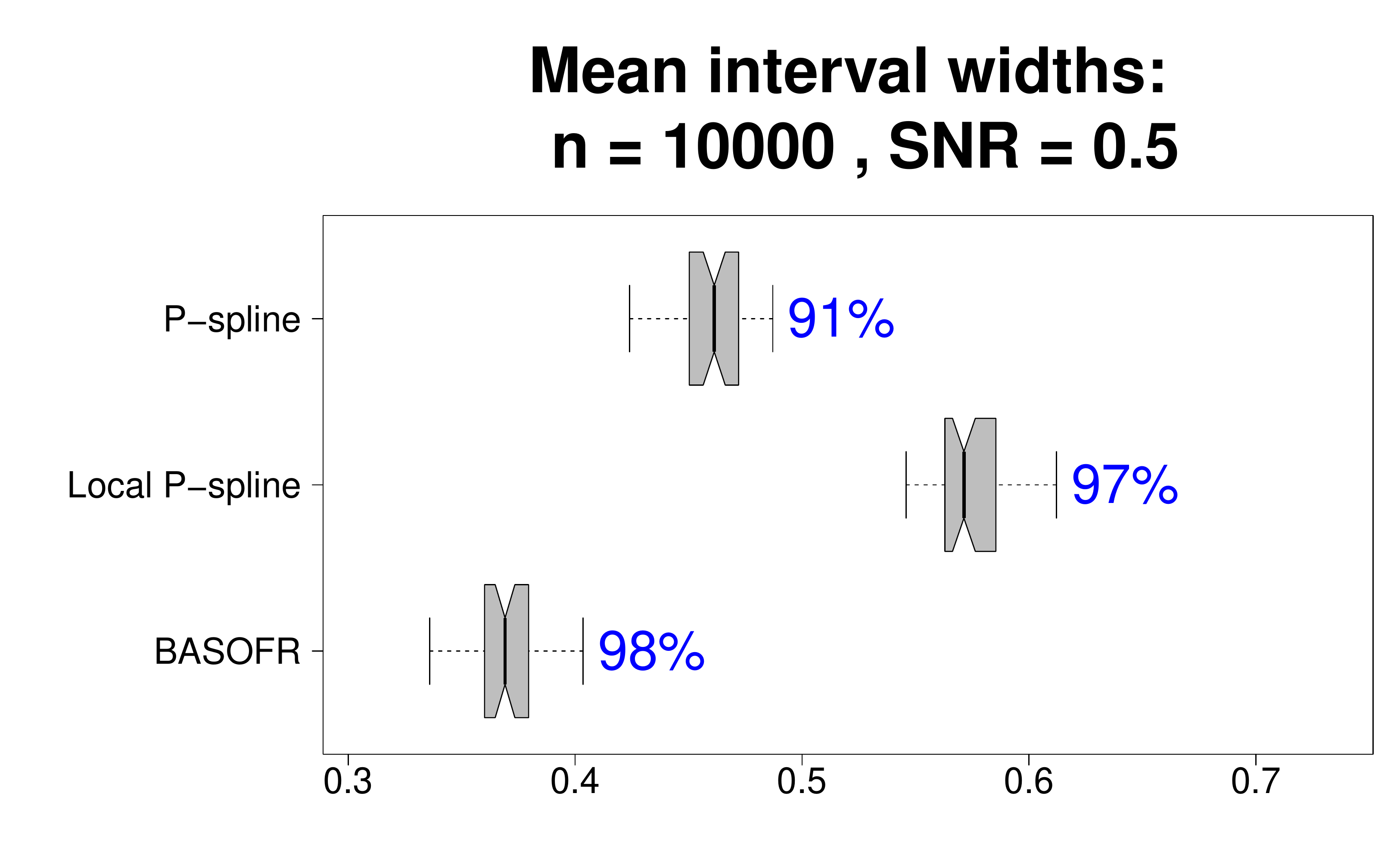}
    \caption{Mean credible interval widths (boxplots) and empirical pointwise coverage  (blue  annotations) for the 95\% credible intervals computed under each model.  BASOFR offers substantially more precise uncertainty quantification that maintains the nominal coverage.} %The $(n, \mbox{SNR}) = (500, 5)$ case is similar to $(n, \mbox{SNR}) = (10000, 0.5)$ and is  omitted.}
    \label{fig:uq_seasonal}
\end{figure}

Point estimation is evaluated using  $L_2$-error  (Figure~\ref{fig:l2error_seasonal}) and uncertainty quantification is evaluated using mean credible interval widths and empirical coverage (Figure~\ref{fig:uq_seasonal}). Most notably, the proposed BASOFR model provides highly accurate point estimates and narrow interval estimates that achieve the nominal coverage, with the most substantial gains over competing methods occurring for  larger sample sizes.   By comparison, the P-spline and local P-spline intervals are far too conservative. Thus, neither global scale parameters nor local but independent scale parameters are sufficient for effective and adaptive inference:  the \emph{dependence} induced by the DHS prior in \eqref{dhs_Bspline} is critical. 
Lastly, the narrow intervals provided by BLISS are far from achieving the nominal coverage and thus inadequate.

%We note that including the seasonality pattern in the functional covariates has made estimating the regression coefficient function more challenging (see supplement). This outcome is expected as the seasonality patterns are introducing additional autocorrelation structure into the functional covariates, and more highly autocorrelated functional covariates usually present more challenges for fitting a SOFR model. 

\subsection{Decision analysis for selecting critical windows}\label{sims_2}
\begin{figure}
    \centering
    \includegraphics[width=\textwidth]{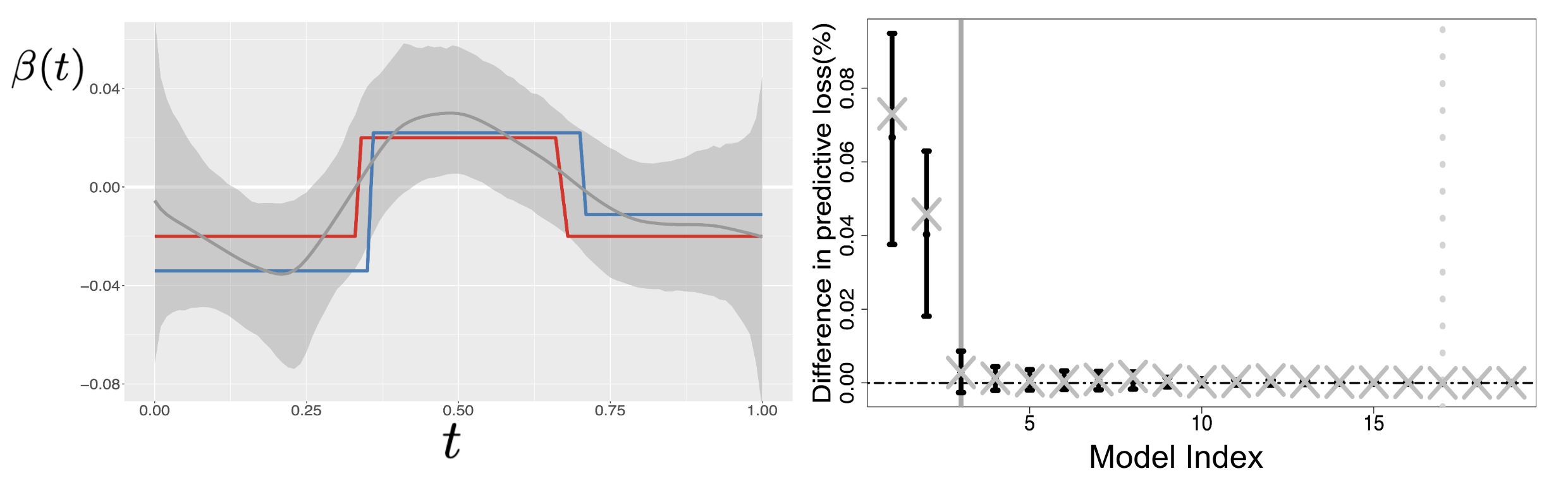}
    \caption{ {\bf Left:} True regression coefficient function (red) with the posterior mean (gray line), 95\% credible interval (shade), and locally constant estimate (blue) for $\beta$ under BASOFR for simulated data. {\bf Right:} for each $\lambda$ in the solution path of \eqref{fusedlasso},  expectations and 95\% credible intervals for the percent difference in predictive mean squared error $\widetilde {\mathcal{D}}_{\lambda}$ (black lines) and the analogous empirical version (x-marks) based on \eqref{empirical}. The vertical lines denote $\lambda_{\min}$ (dotted gray) and the simplest acceptable $\lambda$ (solid gray).}%The \textit{domain selection} property of the proposed posterior summarization tools enables accurate estimation for the regions where the functional predictors that are positively or negatively associated with the outcome variable.}
    \label{DA_example}
\end{figure}

Next, we evaluate whether the proposed decision analysis approach (Section~\ref{PS}) is able to identify critical windows of susceptibility. We simulate 50 datasets from the SOFR model \eqref{SOFR} with $n=100,000$ observations with a low signal-to-noise ratio (SNR $=0.5$) using a locally constant function for the true regression coefficient (see Figure \ref{DA_example}) and the seasonal functional covariates $X_i(t)$. This design is constructed to mimic the output from the real data analysis (see Section~\ref{apps}, Figure~\ref{EOG_PM2.5_v2}), and especially the large sample size and low SNR. 
After fitting the BASOFR model, we apply the proposed decision analysis  and select  $\hat \delta_\lambda$ to be the simplest member of the acceptable family $\mathcal{A}_{0.1}$, i.e. the locally constant approximation with the fewest changes in the local level.

First, we summarize the results on a single simulated dataset in Figure \ref{DA_example}. Most notably, the posterior mean and credible intervals for $\beta$---as summaries of the BASOFR posterior---offer limited ability to describe the true regression function, both in terms of shape and effect direction. However, the decision analysis approach---which is based on the same BASOFR posterior---adequately recovers the truth. Figure~\ref{DA_example} (right panel) also shows that other locally constant estimates are equally competitive, but the selected version is the simplest.

Next, we simulate 50 datasets from the same design and use the same two-stage procedure to fit the BASOFR and extract  $\hat \delta_\lambda$. %Among these locally constant estimates, we identify $\hat \delta_k > 0.005$ as positive and $\hat \delta_k < -0.005$ as negative. 
As a competing method, we select (positive or negative) windows based on whether the 95\% posterior credible intervals for $\beta$ exclude zero. This intuitive and popular strategy is based on the same BASOFR that is used in the decision analysis approach, and thus differs only in the selection criteria. 

\begin{figure}
    \centering
    \includegraphics[width = \textwidth]{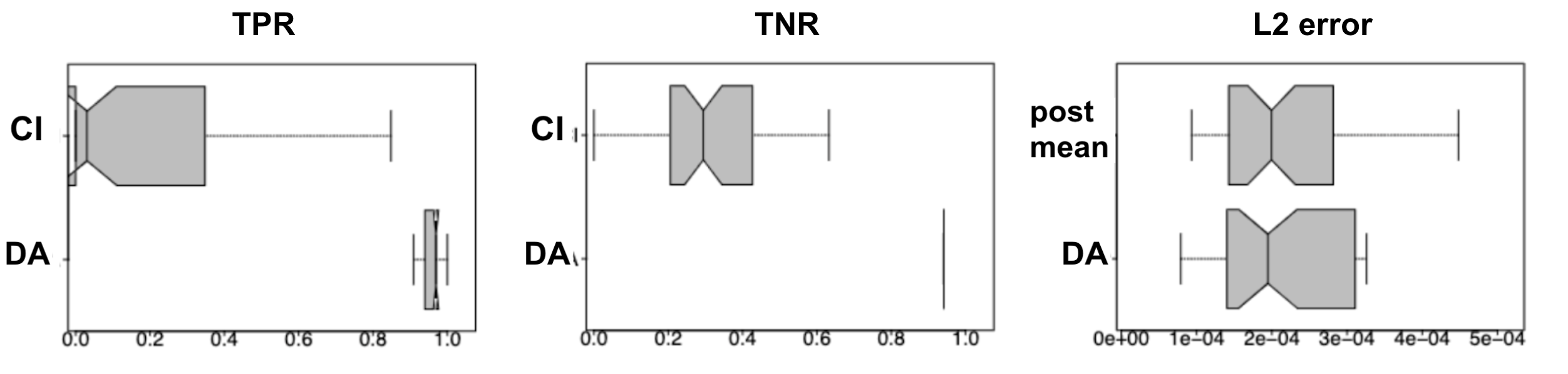}
    \caption{TPRs ({\bf left}) and TNRs ({\bf center}) show significantly better window selection for the proposed decision analysis (DA) approach compared to credible interval-based selection (CI), with no apparent loss in point estimation accuracy compared to the posterior mean ({\bf right}). 
    %to compare window selection using credible intervals (CI) and the proposed decision analysis (DA) approach, along with $L_2$-errors to compare point estimation from $\hat \beta$ and $\hat \delta_\lambda$. The proposed DA approach offers significantly better window selection with no apparent loss in point estimation accuracy.
    }
    \label{fig:DA_sim_plot}
\end{figure}

To evaluate these approaches, we compute the true positive (TPR) and true negative (TNR) rates, defined here to be the correct detection of a truly positive (respectively, negative) window for the regression coefficient function $\beta$ (Figure~\ref{fig:DA_sim_plot}).  Most notably, the proposed decision analysis approach is significantly better at selecting the critical windows of susceptibility, while selection based on credible intervals of $\beta$ is far too conservative and thus underpowered. We emphasize that this important result applies for the BASOFR posterior credible intervals, which are substantially tighter (with the correct coverage) than competing interval estimates (Figure~\ref{fig:uq_seasonal}). Thus, alternative Bayesian SOFR models with less precise (wider) interval estimates would offer even less power to select these critical windows. We also compute the $L_2$-error of the point estimates from $\hat \delta_\lambda$ and $\hat \beta$, which confirms that the 
locally constant estimator does not sacrifice point estimation accuracy compared to the posterior mean.

%Additionally, even though the locally constant estimator is much simpler than the posterior mean (see Figure~\ref{DA_example}), there is no loss in estimation accuracy, which achieves the goal of the acceptable family \eqref{accept_family}.

%The $L_2$-errors are similar, which is by design: $\hat \delta_k$ is optimized to match or nearly match the predictive performance. 

 %Notably, the 95\% credible intervals are far too conservative under the low signal-to-noise ratio scenario, which was previously noted in Figure \ref{DA_example}. Hence, simply checking whether the credible intervals exclude zero or not is not sufficient to detect the critical windows: the \textit{domain selection} provided by the proposed posterior summarization tools greatly improves the critical window identification results compared to the posterior credible bands among both positive and negative critical windows and with no apparent loss at the $L_2$-error compared to the posterior expectation. 

The  key takeaway from Figures~\ref{DA_example}--\ref{fig:DA_sim_plot} is that, despite using a smooth B-spline basis  for $\beta$, the BASOFR posterior distribution \emph{does} contain enough information to estimate a (true) locally constant regression function, but requires a careful decision analysis (Section~\ref{PS})---rather than traditional posterior summaries (posteriors means and credible intervals)---to access it.

\section{Prenatal $\mbox{PM}_{2.5}$ exposure and educational outcomes}\label{apps}
We apply the BASOFR model and accompanying decision analysis to study the effects of prenatal exposure to $\mbox{PM}_{2.5}$ on educational outcomes. Specifically, we deploy the SOFR model \eqref{SOFR} for standardized 4th EOG reading score $y_i \in \mathbb{R}$, $\mbox{PM}_{2.5}$ exposure during gestation $X_i\!:\mathcal{T}_i \rightarrow \mathbb{R}$, and other scalar covariates $\bm z_i \in \mathbb{R}^p$  (see Table~\ref{tab:variable}) for a large cohort of mother-child pairs   $i=1,\ldots,n$ in NC. 

The use of model \eqref{SOFR} requires careful consideration of the domains $\mathcal{T}_i$ and the scalar covariates $\bm z_i$. First,  each domain is subject-specific: $\mathcal{T}_i = [1, T_i]$, where $T_i$ is the number of days in the gestational period for mother-child pair $i$. The gestation lengths range from 30 to 42 weeks, so the total domain is $\mathcal{T} = [1,T_{\max}]$, where $T_{\max} = 295$ days is the longest gestational period in the dataset. The B-spline basis is defined on this interval. 

 Next, the covariates $\bm z_i$ are given  in Table~\ref{tab:variable} (with the exception of  \texttt{Reading\_Score} and \texttt{Prenatal$\mbox{PM}_{2.5}$}). Each continuous covariate is centered and scaled, and the categorical variables 
 %\texttt{mEdu}, \texttt{mRace} and \texttt{BirthMonth} 
 are encoded using dummy variables. %, with reference groups \texttt{HS}, \texttt{White}, and no reference group, respectively. 
We modify \eqref{SOFR} to include nonlinear additive effects for mother's age (\texttt{mAge}), length of gestation $T_i$ (\texttt{Gestation}), and age-within-cohort (\texttt{Age\_w\_cohort}). The length of gestation is not only biologically important---and potentially nonlinear---but also $T_i$ appears in the key functional term $\int_{\mathcal{T}_i} X_i( t)\beta (t)\ dt$ in \eqref{SOFR}, so a flexible accounting for the effect of $T_i$ is crucial. For \texttt{mAge} and \texttt{Gestation}, we use piecewise continuous linear splines with knots at  ages 18, 24, 29, 34 and  weeks 34, 37, 39, 41, respectively. The coefficients corresponding to these linear and nonlinear effects   (except \texttt{Age\_w\_cohort}, see below) are  assigned the hierarchical prior $[\alpha_j \mid \sigma_j] \sim \mathcal{N}(0,\sigma_j^2)$ with $\sigma_j^{-2} \sim \text{Gamma}(0.01,0.01)$ to encourage shrinkage and guard against the effects of multicollinearity among the correlated covariates $\bm z_i$ (see the supplementary material).

For  age-within-cohort, we anticipate that older students may perform better on their standardized tests, but only up to a point: students who are more than one year older than their classmates may have repeated a grade or enrolled in kindergarten later for developmental reasons. Thus, we model \texttt{Age\_w\_cohort} as a nonlinear effect, and in particular use the proposed adaptive B-spline model with dynamic shrinkage processes \eqref{b-spline sofr}--\eqref{DHS} for this term (as well as $\beta$ in \eqref{SOFR}). This specification encourages the nonlinear effect of \texttt{Age\_w\_cohort} to be smooth, but can capture rapid changes  such as those expected around 52 weeks. Additional details and summary statistics for \texttt{Age\_w\_cohort} are in the supplementary material.

Posterior inference from the BASOFR model is based on 10,000 draws from the Gibbs sampler (after discarding a burn-in of 10,000). Traceplots show no lack of convergence and effective sample sizes are sufficiently large.

\begin{figure}
    \centering
    \includegraphics[width=\textwidth]{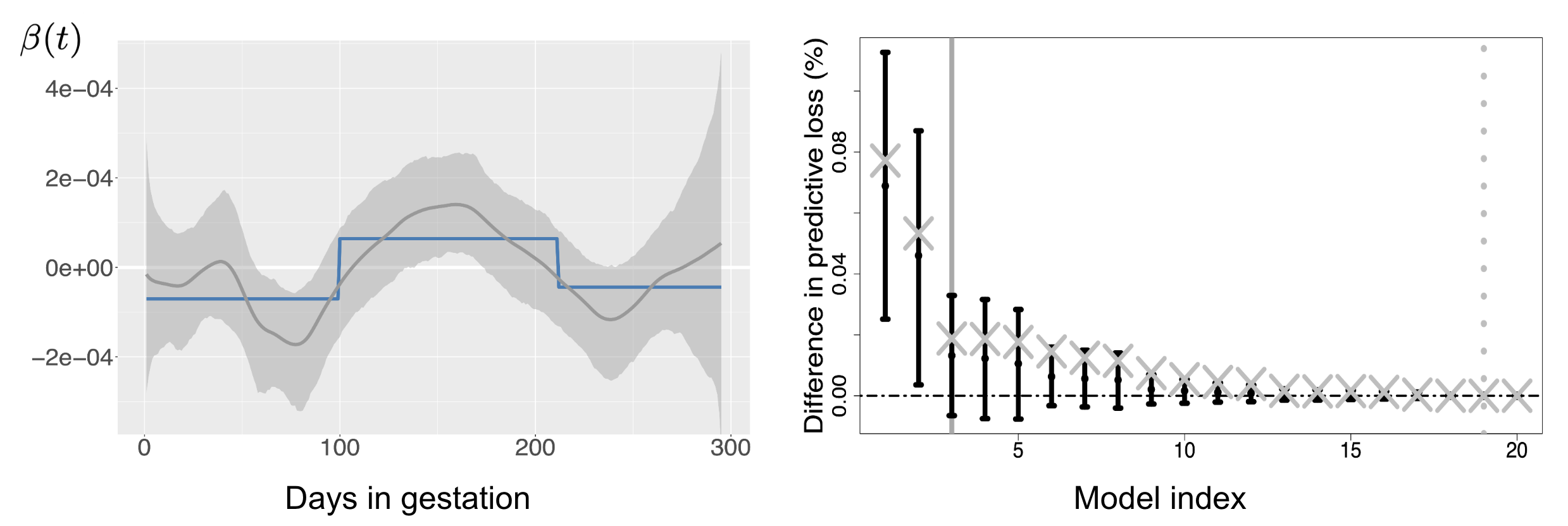}
    \caption{{\bf Left:} Posterior mean (gray line), 95\% credible interval (shade), and locally constant estimate (blue) for $\beta$ under BASOFR for the NC data. {\bf Right:}   Expectations and 95\% credible intervals for the percent difference in predictive mean squared error $\widetilde {\mathcal{D}}_{\lambda}$ (black lines) and the analogous empirical version (x-marks)  based on \eqref{empirical} for each $\lambda$ in the solution path of \eqref{fusedlasso}. The vertical lines denote $\lambda_{\min}$ (dotted gray) and the simplest acceptable $\lambda$ (solid gray).}
    \label{EOG_PM2.5_v2}
\end{figure}

First, we summarize our inference on the regression coefficient function $\beta$ in Figure~\ref{EOG_PM2.5_v2}, which includes  traditional posterior summaries of $\beta$ (posterior means and 95\% credible intervals) along with the proposed locally constant point estimate. 
We select $\hat \delta_\lambda$ to be the simplest member  (i.e., the locally constant estimate with the fewest changes in the local level)  of the acceptable family $\mathcal{A}_{0.1}$. Figure~\ref{EOG_PM2.5_v2} (right panel) justifies this choice: the simplest member of the acceptable family indeed provides near-optimal prediction compared to the other point estimators along the solution path of \eqref{fusedlasso}. Notably, the locally constant estimator substantially simplifies the shape of $\beta$ and selects the critical windows of susceptibility. We refer the three locally constant regions in $\hat\delta_\lambda$ as R1, R2 and R3, which are similar but not identical to trimesters one, two, and three, respectively.

\begin{figure}
    \centering
    \includegraphics[width=\textwidth]{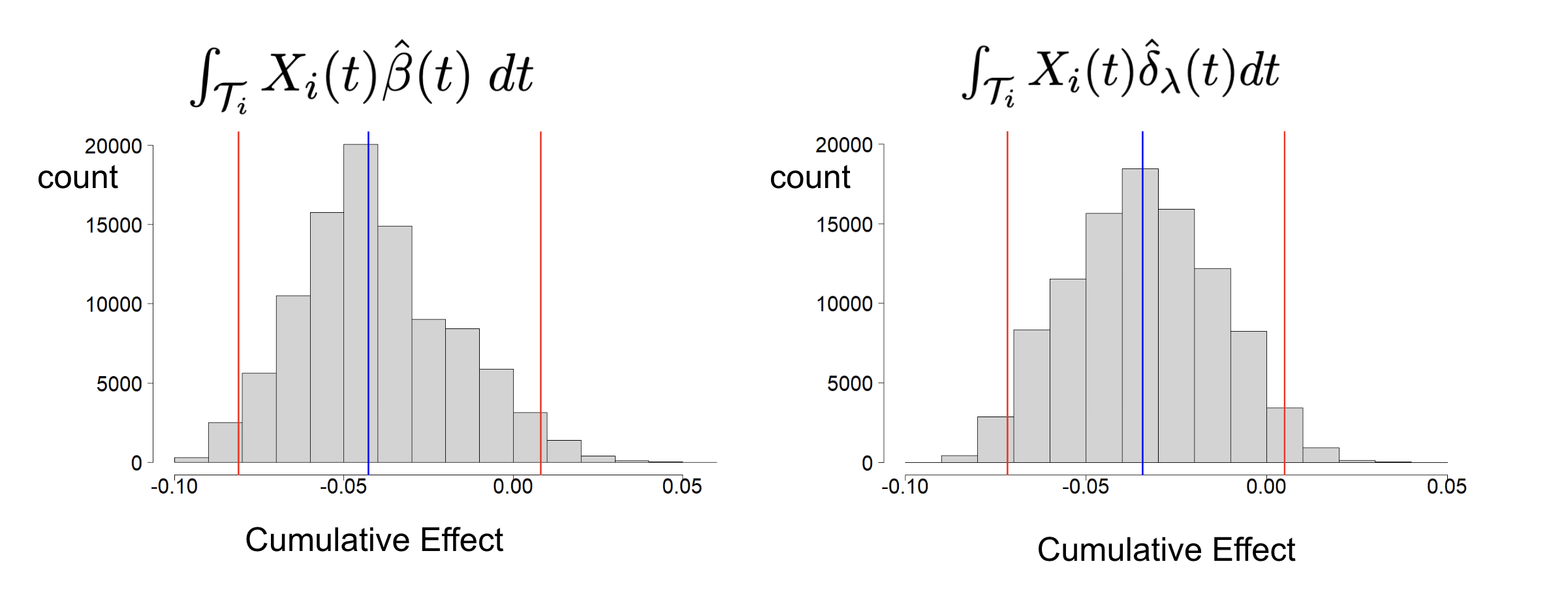}
    \caption{The estimated expected cumulative effects of $\mbox{PM}_{2.5}$ exposure during pregnancy across mother-child pairs $i=1,\ldots,n$ using the  posterior mean $\hat\beta$ ({\bf left}) and $\hat \delta_\lambda$ ({\bf right}). The sample median (blue) and 95\% intervals (via sample quantiles; red) are annotated. The estimated cumulative effects agree and are significantly negative for nearly all mother-child pairs.}
    \label{fig:est-effects}
\end{figure}

Despite the simplifications offered by the decision analysis, the interpretation of the regression coefficient function estimates requires some care. At first glance, Figure~\ref{EOG_PM2.5_v2} suggests that $\mbox{PM}_{2.5}$ exposure is detrimental in R1 and R3 yet \emph{favorable} in R2. Such a contradictory effect seems implausible. To investigate this outcome, we compute the estimated cumulative effect of  exposure to $\mbox{PM}_{2.5}$,  $ \int_{\mathcal{T}_i} X_i( t)\hat \beta (t)\ dt$,  for each mother-child pair $i=1,\ldots,n$ (Figure~\ref{fig:est-effects}), using both the posterior mean and the locally constant point estimator for $\beta$. The cumulative effect of $\mbox{PM}_{2.5}$ exposure during gestation is \emph{significantly negative} for nearly all mother-child pairs. Thus, as anticipated, exposure to  $\mbox{PM}_{2.5}$ during gestation is negatively associated with 4th EOG reading scores (adjusting for $\bm z_i$).

\begin{figure}
    \centering
    \begin{subfigure}[b]{0.33\textwidth}
         \centering
         \includegraphics[width=\textwidth]{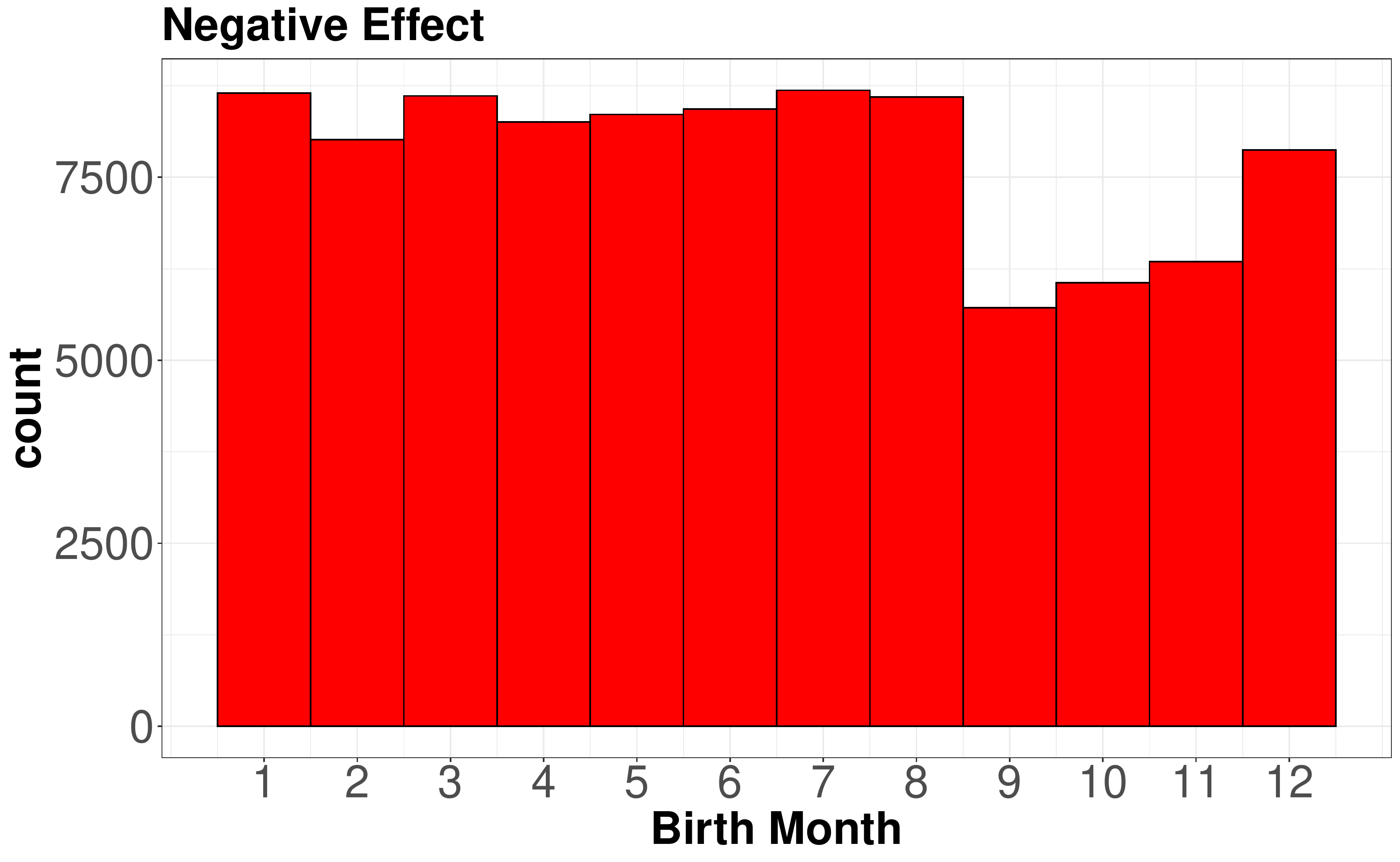}
         
                  \includegraphics[width=\textwidth]{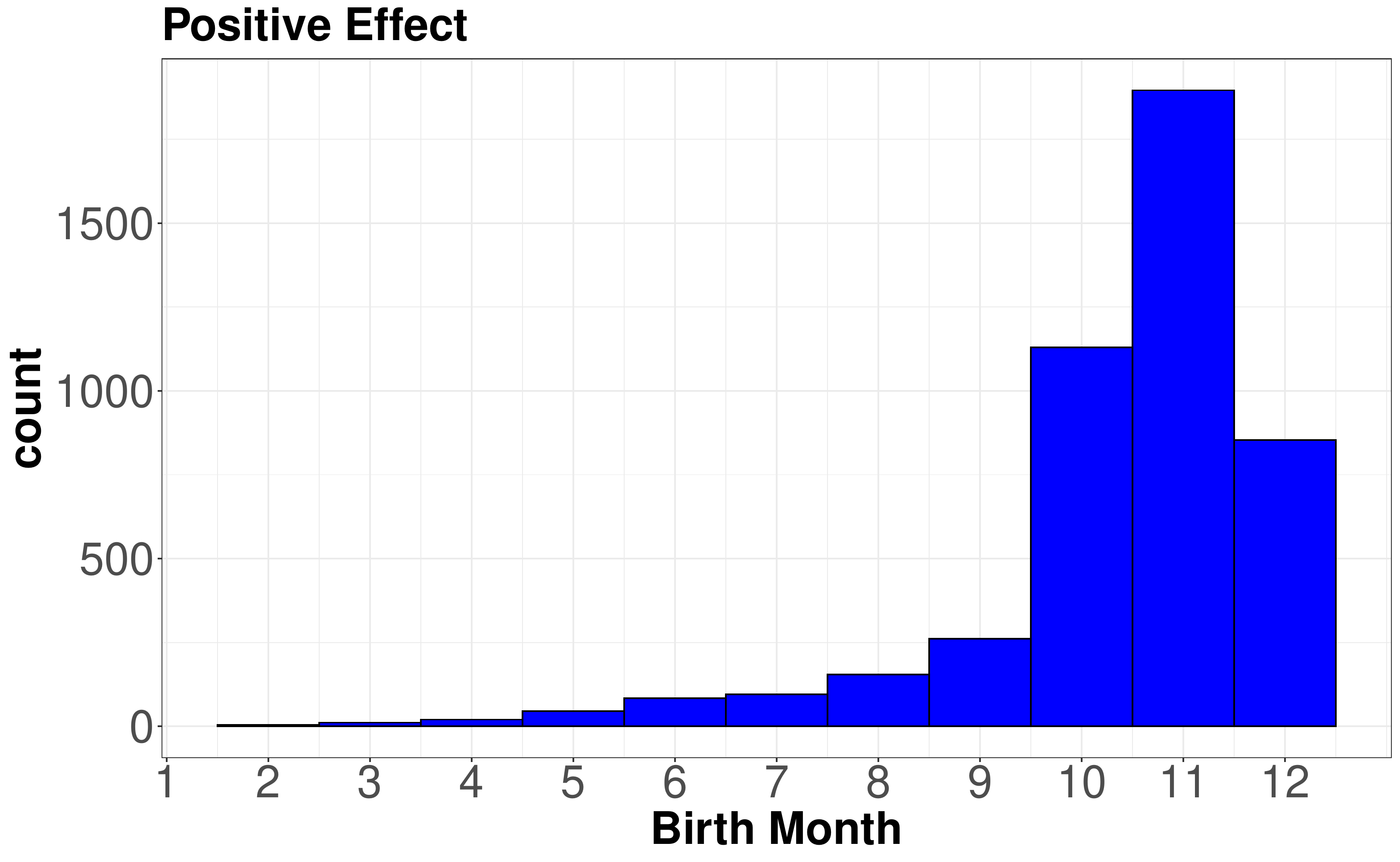}
     \end{subfigure}
     \begin{subfigure}[b]{0.65\textwidth}
         \centering
             \includegraphics[width = \textwidth]{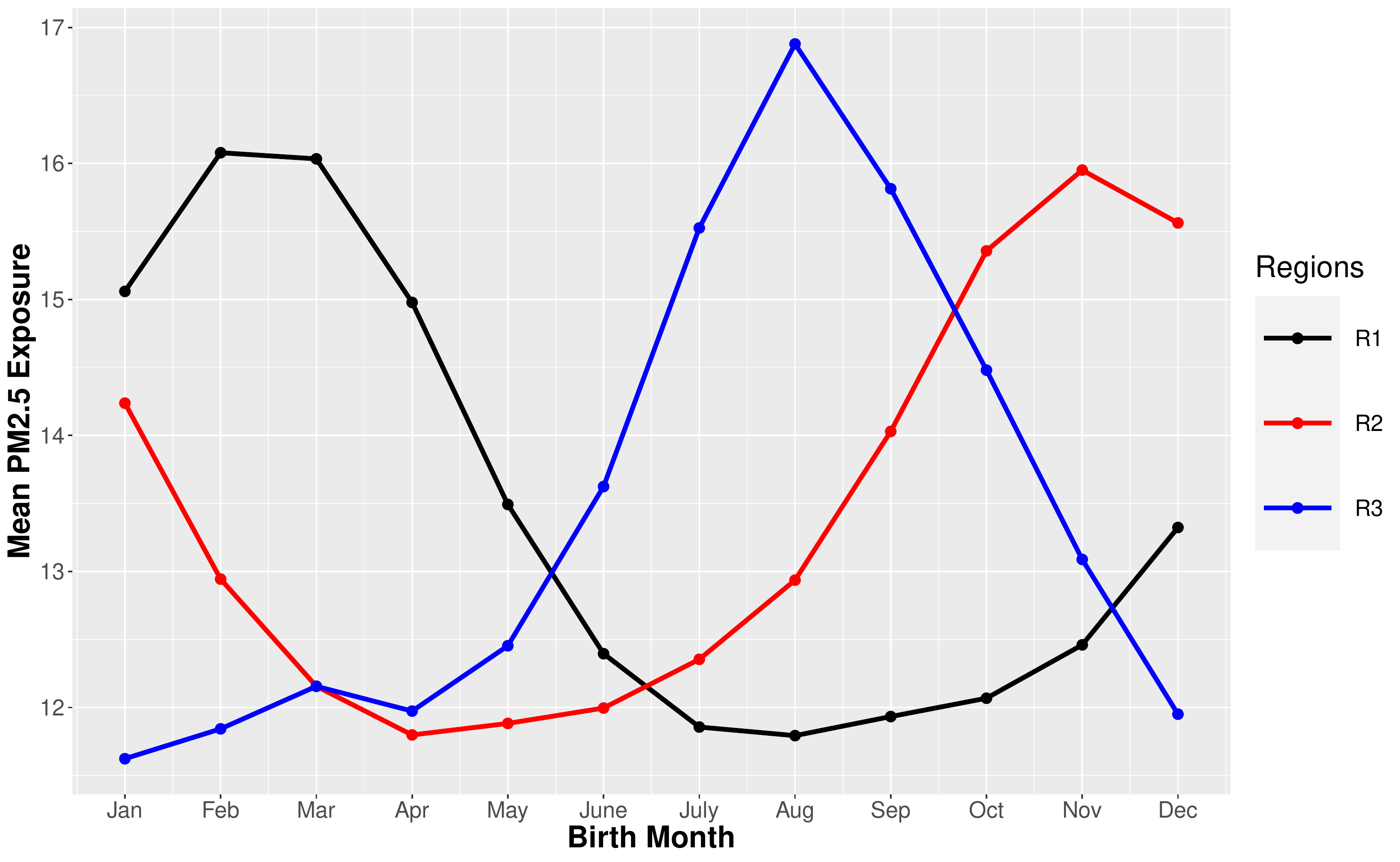}
     \end{subfigure}
    \caption{Distribution of the birth month for mother-child pairs with a negative ({\bf top left}) or positive ({\bf bottom left}) estimated expected cumulative effect   of $\mbox{PM}_{2.5}$ exposure during pregnancy, along with the average $\mbox{PM}_{2.5}$ exposures over the windows identified by $\hat \delta_\lambda$ (see Figure~\ref{fig:est-effects}). The positive effects represent a small fraction of mother-child pairs with birth months almost exclusively in October-December, which corresponds to high exposures during R2 but low exposures during R1 and R3. This seasonality is vital for interpreting $\beta$.}
    \label{fig:seasonal}
\end{figure}

We further investigate \emph{which} students were assigned positive or negative estimated cumulative effects $ \int_{\mathcal{T}_i} X_i( t)\hat \beta (t)\ dt$ (Figure~\ref{fig:seasonal}). Notably, the students for which this effect is (unexpectedly) positive were born almost exclusively in October-December. This result corresponds to a seasonal pattern in daily $\mbox{PM}_{2.5}$ exposure, which is further confirmed in Figure \ref{fig:seasonal} (right panel): the birth month determines the  average $\mbox{PM}_{2.5}$ exposure over each region R1, R2, and R3, with October-December corresponding uniquely to high exposures during R2 but low exposures during R1 and R3. Thus, the positive estimate in R2 is confounded by low exposures during R1 and R3 (Figure~\ref{EOG_PM2.5_v2}). We emphasize that seasonality is already included in the model via $\bm z_i$: both birth month and age-within-cohort are included as nonlinear effects, and capture overlapping yet mutually important notions of seasonality. 

Our cumulative analysis---the estimated effects and windows selected  (Figure~\ref{EOG_PM2.5_v2}), the overwhelmingly negative cumulative effects across mother-child pairs (Figure~\ref{fig:est-effects}), and the seasonality patterns  (Figure~\ref{fig:seasonal})---leads us to conclude that R1 and R3 represent the critical windows of susceptibility that are adversely associated with 4th EOG reading scores.

\begin{figure}
    \centering
    \includegraphics[width=0.49\textwidth]{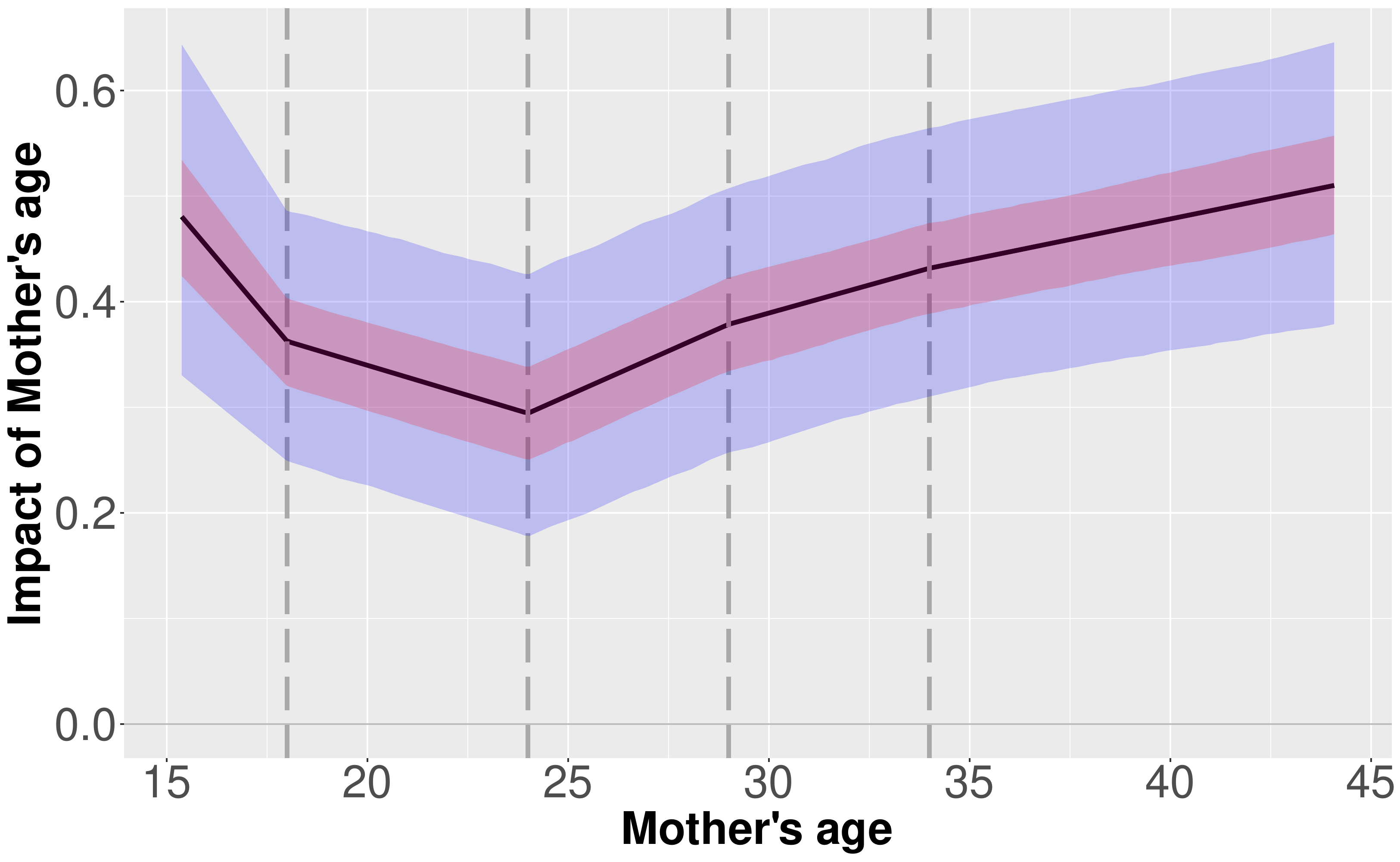}
    \includegraphics[width=0.49\textwidth]{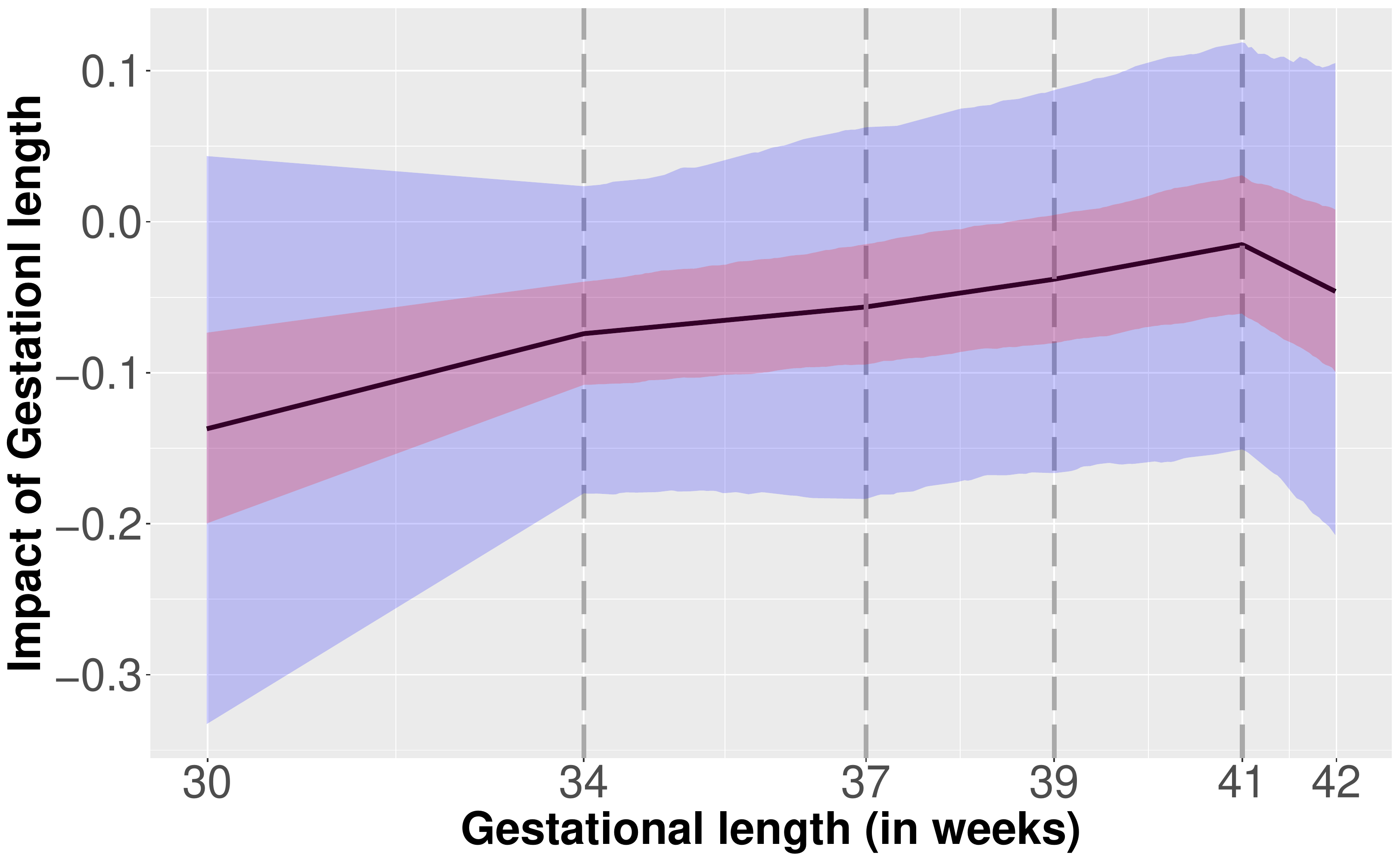} 

    \includegraphics[width=0.49\textwidth]{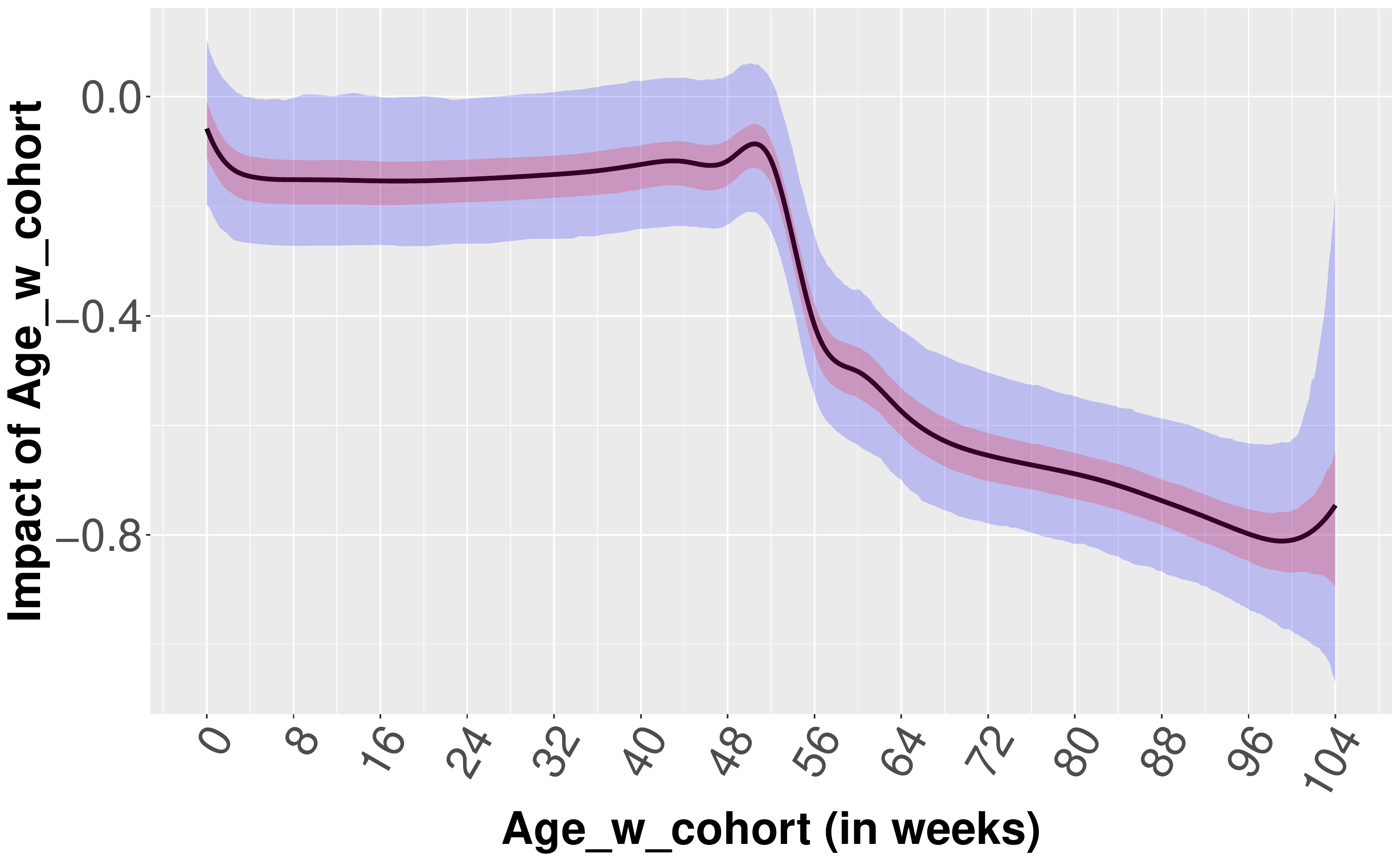}
    \includegraphics[width=0.49\textwidth]{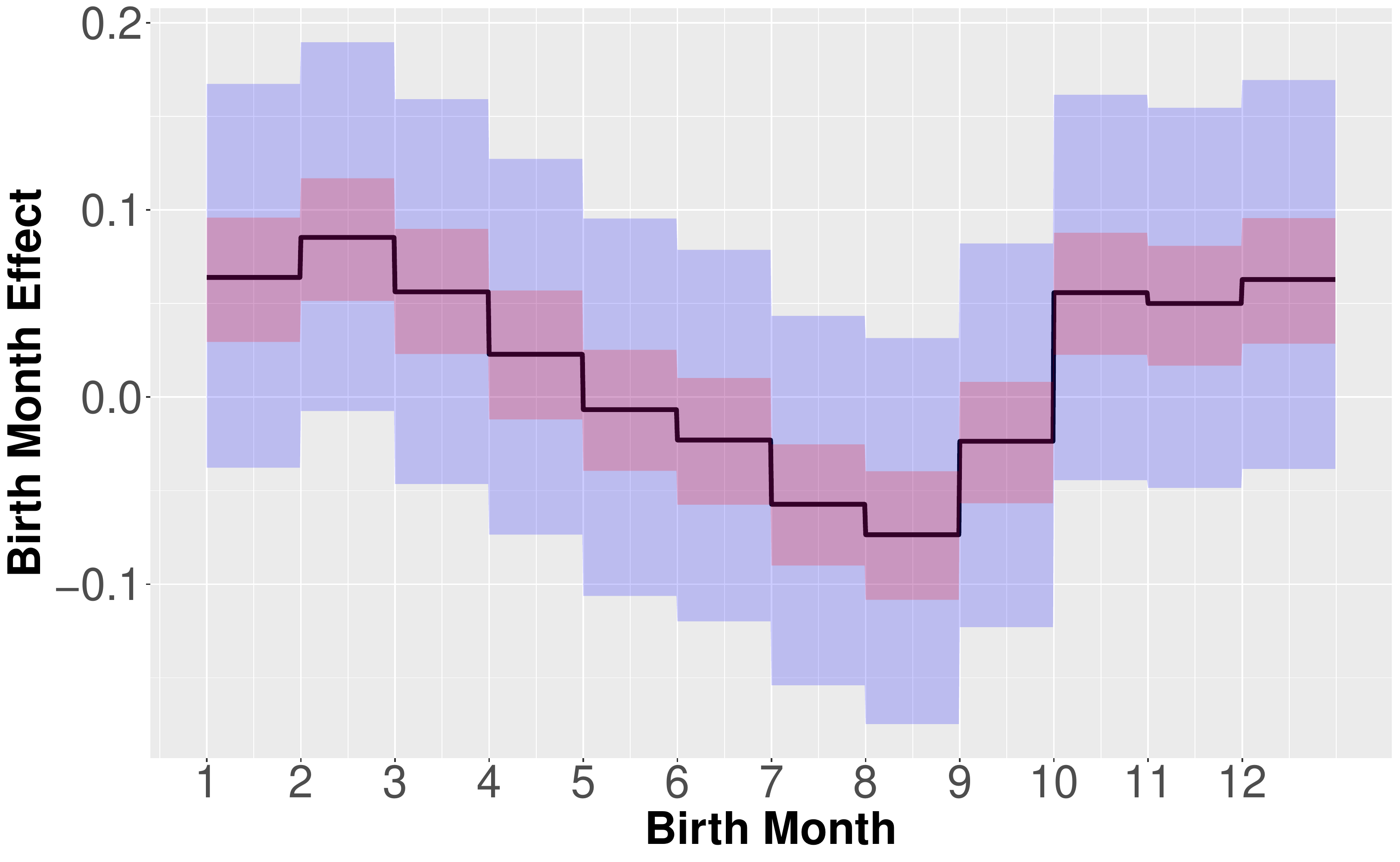} 
    \caption{The posterior mean (line) and 50\% (red) and 95\% (blue) credible intervals for the nonlinear effects of mother's age at birth ({\bf top left}), gestational length ({\bf top right}), age-within-cohort ({\bf bottom left}) and birth month ({\bf bottom right}) on reading scores.}
    \label{fig:nonlinear}
\end{figure}

Lastly, we summarize the posterior inference for the scalar covariates $\bm z_i$. Among the linear effects (Table \ref{reg_result_2}), we find that lower 4th EOG reading scores are associated with lower mother's education level, presence of  economic disadvantages, higher blood lead levels, smoking, and race/ethnicity and gender. The nonlinear effects are presented in Figure~\ref{fig:nonlinear}. Mother's age is positively associated with higher 4th EOG reading scores after age 24. The negative association prior to age 24 is perhaps explained by strong correlations between this younger age group and (lower) mother's education levels (see the supplement), which is already strongly associated with $y_i$. Gestational length is positively associated with 4th EOG reading scores until about 41 weeks, at which point the pregnancy is considered late term and accompanied by other health complications. Finally, age-within-cohort and birth month are highly correlated, and thus these effects must be interpreted jointly.  The larger effects for birth months October-March are likely explained in part because those students are typically older within their cohort, which further explains why the age-within-cohort effect has only a small positive slope prior to week 52. However, for students at least one year older than their cohort, the age-within-cohort effect is significantly negative and includes rapid changes in the regression function---which justifies the choice of the adaptive B-spline model with dynamic shrinkage processes \eqref{b-spline sofr}--\eqref{DHS}.

\begin{table}
\centering 
\caption{Posterior means and 95\% credible intervals for the scalar regression coefficients $\bm \alpha$.  Intervals that exclude zero are annotated (**).}
\begin{tabular}{llll}
\rowcolor[HTML]{C0C0C0} 
Covariate                                             & \begin{tabular}[c]{@{}l@{}}Regression \\ Coefficient Estimates\end{tabular}                                    & Covariate                                                   & \begin{tabular}[c]{@{}l@{}}Regression \\ Coefficient Estimates\end{tabular}                 \\
\multicolumn{1}{l|}{{noHS}}                   & \multicolumn{1}{l|}{\begin{tabular}[c]{@{}l@{}}-0.14  (-0.16, -0.13)**\end{tabular}}                         & \multicolumn{1}{l|}{Male}                                   & \begin{tabular}[c]{@{}l@{}}-0.13  (-0.14, -0.12)**\end{tabular}                           \\
\rowcolor[HTML]{EFEFEF} 
\multicolumn{1}{l|}{\cellcolor[HTML]{EFEFEF}higherHS} & \multicolumn{1}{l|}{\cellcolor[HTML]{EFEFEF}\begin{tabular}[c]{@{}l@{}}0.28 (0.26, 0.29)**\end{tabular}}    & \multicolumn{1}{l|}{\cellcolor[HTML]{EFEFEF}EconDisadvantage} & \cellcolor[HTML]{EFEFEF}\begin{tabular}[c]{@{}l@{}}-0.27 (-0.29, -0.26)**\end{tabular}   \\
\multicolumn{1}{l|}{NH Black}                         & \multicolumn{1}{l|}{\begin{tabular}[c]{@{}l@{}}-0.49 (-0.51, -0.48)**\end{tabular}}                         & \multicolumn{1}{l|}{Smoker}                                 & \begin{tabular}[c]{@{}l@{}}-0.07 (-0.09, -0.06)**\end{tabular}                           \\
\rowcolor[HTML]{EFEFEF} 
\multicolumn{1}{l|}{\cellcolor[HTML]{EFEFEF}Hispanic} & \multicolumn{1}{l|}{\cellcolor[HTML]{EFEFEF}\begin{tabular}[c]{@{}l@{}}-0.06  (-0.08, -0.04)**\end{tabular}} & \multicolumn{1}{l|}{\cellcolor[HTML]{EFEFEF}Blood\_level}   & \cellcolor[HTML]{EFEFEF}\begin{tabular}[c]{@{}l@{}}-0.028  (-0.033, 0.022)**\end{tabular}
\end{tabular}
\label{reg_result_2}
\end{table}

\section{Conclusion}\label{conclude}
We developed a fully Bayesian modeling, computational, and decision analysis framework to study the effects of prenatal exposure to air pollution on educational outcomes. The proposed Bayesian adaptive scalar-on-function regression (BASOFR) model was designed to capture both smooth and abrupt changes in the association function, provide full posterior uncertainty quantification, and maintain computational scalability in both the sample size and the number of observation points for each functional covariate. To complement the BASOFR model, we developed a decision analysis  approach that produces locally constant point estimates of the regression coefficient function, which simultaneously (i) extracts the critical windows of the functional domain, (ii) partially   resolves fundamental interpretability issues for SOFR, and (iii) only adds minimal computational cost. Simulation studies demonstrated substantial advantages of the proposed approach for point estimation,  uncertainty quantification, window selection, and computational scalability. 

Using the proposed approach, we analyzed a large cohort ($n \approx 100,000$) of mother-child pairs in North Carolina to study the effects of $\mbox{PM}_{2.5}$ exposure during gestation on  4th end-of-grade reading scores. This analysis required careful consideration of the SOFR model output---including effect directions, cumulative effects, and seasonality---to identify two critical windows of susceptibility around trimesters one and three that correspond to adverse educational outcomes. Crucially, these results were enabled by our joint consideration of adaptive regression modeling, computational scalability, and interpretable posterior summaries via customized decision analysis. Our analysis included adjustments for important confounding variables and estimated nonlinear regression effects for mother's age, gestation length, birth month, and age-within-cohort.

We note that the estimated regression coefficient function (Figure~\ref{EOG_PM2.5_v2}), while complex, does not exhibit the same rapidly-changing features as in other examples (Figures~\ref{quick_example}, \ref{DA_example}, and \ref{fig:nonlinear}). However, the use of BASOFR remains justified: (i) the function shape was not known in advance, so the more flexible modeling capability is reassuring, and  (ii) our simulation studies decisively showed that BASOFR delivers more accurate estimates and more precise and well-calibrated uncertainty quantification than competing methods. Regardless, the proposed Bayesian specification for adaptive regression coefficient functions was highly useful for inferring the nonlinear age-within-cohort effect, which exhibited both smooth and rapidly-changing features (Figure~\ref{fig:nonlinear}). Thus, our adaptive Bayesian approach is useful not only in SOFR, but in function estimation and additive models more broadly.

The proposed functional regression model and accompanying decision analysis strategies offer promising extensions. First, these methods may be generalized for multiple functional predictors as well as functional response variables in function-on-function regression. Second, our decision analysis approach is broadly applicable for Bayesian SOFR and related distributed lag models, and thus is a useful addition to these models for more powerful window selection and interpretable model summaries. Furthermore, the decision analysis strategy may be altered to induce other structured point estimates, such as sparse or locally linear summaries, by varying the penalty term in \eqref{loss_locally_constant}. Lastly, our procedures may be applied to other datasets to estimate the critical windows of susceptibility for various exposures or interventions measured at high resolutions and paired with important outcomes of interest. 

\clearpage

\section*{Supplement}

\appendix

The supplement contains the following: additional details about the computing (Section~\ref{comp}), additional simulation results (Section~\ref{add_sims}), and additional information and summary statistics for the North Carolina (NC) dataset (Section~\ref{add_apps}).

\section{Computing Details}\label{comp}
This section describes the functional data preprocessing and the Markov chain Monte Carlo (MCMC) algorithm for the proposed Bayesian adaptive scalar-on-function regression (BASOFR):
\begin{align}
   y_i =\mu + \bm {z}_i' \bm {\alpha} + \int_{\mathcal{T}_i}X_i(t)\beta(t)\ dt + \epsilon_i, \quad [\epsilon_i \mid \sigma^2] \stackrel{\text{iid}}{\sim} \mathcal{N}(0, \sigma^2), \quad i=1,\ldots,n \label{sofr} 
\end{align}
using the priors given in the main paper.

We first show how we preprocess the functional data. Let $\bm x_i = (x_{i,1},\ldots, x_{i,m_i})'$ denote the observation of the $i$th function $X_i$ (i.e., $\mbox{PM}_{2.5}$ exposure profile for student $i$) at $m_i$ points. We convert this to a functional predictor $X_i$ using a B-spline basis expansion $X_i(t) = \sum_{k=1}^{K_X} X_{ik}^* \phi_k(t)$ with the coefficients $\{X_{ik}^*\}$ estimated using ordinary least squares; other estimates may be used. This step smooths over the noisy observations and resolves the difficulties of irregular observation points.
%; however, it is not strictly necessary for the BASOFR model. 
Next, we expand $\beta(t)=\sum_{k=1}^{K_B} B_k^* \psi_k(t)$ for a known equally spaced B-spline basis $\{\psi_k\}_{k=1}^{K_B}$. In our simulation study and application study, we expand the $\{X_i\}$ and $\beta$ on the same equally-spaced B-spline basis ($K_X = K_B =53$ in the simulation; $K_X=K_B=103$ in the application study); however, we note that the B-spline basis for the regression coefficient function $\beta$ need not be the same as the basis for $\{X_i\}$. After the basis expansion on $\{X_i\}$ and $\beta$, we now are able to simplify \eqref{sofr} as:
\begin{align}
    y_i  = \mu + \bm z_i' \bm \alpha + \bm X_i^{**} \bm B^{**} + \epsilon_i, \quad [\epsilon_i \mid \sigma^2] \stackrel{\text{iid}}{\sim} \mathcal{N}(0, \sigma^2), \quad i=1,\ldots,n \label{sofr_simple} 
\end{align}
where $\mathbf {X}_i^{**} = \mathbf {X}_i^* \mathbf {J}_i^{\phi,\psi}$ with $\bm X_i^* = (X_{i1}^*, \ldots,X_{iK_X}^*)$ and $\mathbf {J}_i^{\phi, \psi} = [\int_{\mathcal{T}_i}\phi_j(t)\psi_k(t) \ dt]_{jk}$.

We construct our MCMC algorithm for the BASOFR method based on \eqref{sofr_simple}. Specifically, we build an efficient Gibbs sampler composed of the following blocks: (i) the regression coefficient function $\beta$, which is updated via the basis coefficients  $\bm B^*$; (ii) the intercept $\mu$ and regression coefficients   $\bm \alpha$; (iii) the local scale parameters $\{\lambda_k\}_{k=1}^{K_B}$, which are updated via the log-volatilities $h_k = \log\lambda_k^2$ for $k=2,\ldots,K_B-1$ and the boundary terms $\lambda_1 = \lambda_{K_B} = \lambda_0$; (iv) the accompanying log-volatility autoregressive parameters $\{\mu_h, \phi\}$; and (v) the variance components $\sigma^2$ and $\{\sigma_j^2\}$, where the latter variances are the prior variances for $[\alpha_j \mid \sigma_j] \sim \mathcal{N}(0, \sigma_j^2)$. The case of nonlinear and additive regression terms for $\bm z$ is handled subsequently.

%where $\bm \alpha$ are the regression coefficients for the scalar covariates, and we assume a hierarchical prior $[\alpha_j|\sigma_j] \sim \mathcal{N}(0,\sigma_j^2)$ with $\sigma^{-2} \sim \text{Gamma}(0.01,0.01)$. Note that \eqref{sofr} does not include any nonlinear relationship between the scalar covariates $\bm z_i$ and the response variable $y_i$, whereas our application study does include nonlinear relationships.  We present a generalization of the MCMC algorithm to include the piecewise linear associations (i.e. the associations between the EOG reading score and the mother's age and gestational length variables) and the nonparametric association (i.e. the nonlinear association between the EOG reading score and age-within-cohort) used in our application study at the end of this section. 

Combining \eqref{sofr_simple} with the dynamic horseshoe (DHS) priors on the second-differenced basis coefficients $\{\Delta^2 B_k^*\}$ as well as the prior on the B-spline coefficients at boundaries ($B_1^*,B_{K_B}^*$), we obtain a $K_B$-dimensional Gaussian full conditional distribution  $[\bm B^* \mid\cdots] \sim \mathcal {N} (\bm Q_{\bm B^*}^{-1}\bm l_{\bm B^*}, \bm Q_{\bm B^*}^{-1})$ with
\begin{align}
     \bm Q_{\bm B^*} = \sigma^{-2}\bm {X^{**}}{'}&{\bm X^{**}} + \bm D'\bm {\Lambda}^{-1} \bm D,\quad \bm l_{\bm B^*} = \sigma^{-2}\bm {X^{**}}{'} \bm y_c,
\end{align}
where 
\[ \bm D =\left[ \begin{array}{ccccccc}
1 & 0 & 0&\cdots&\cdots&\cdots&0 \\
1 & -2 &1 &0 &\cdots &\cdots&0\\
0& 1 & -2&1 &0&\cdots &0 \\
\vdots&&\ddots&\ddots&\ddots&&\vdots\\
0 & \cdots  &   0& 1&    -2&   1&0\\
0&\cdots   &\cdots & 0&1 &    -2&1\\
0&\cdots&\cdots&\cdots&0&0&1
\end{array} \right],
\]
is a $K_B \times K_B$ second-differencing matrix, 
$\bm y_c \coloneqq (y_1- \mu - \bm z_1'\bm \alpha, y_2-\mu - \bm z_2'\bm \alpha,\ldots, y_n-\mu- \bm z_n'\bm \alpha)^{'}$ is the $n$-dimensional vector of centered observations, 
and $\bm \Lambda\coloneqq$ diag$\{\lambda_k^2\}_{k=1}^{K_B}$ is the diagonal matrix of the prior variances on $B_1^*, \{\Delta^2 B_k^*\}_{k=2}^{K-1}$ and $B_{K_B}^*$. This Gaussian full conditional distribution on $\bm B^*$ allows us to jointly sample the B-spline coefficients $\{B_k^*\}_{k=1}^{K_B}$. The regression function $\beta$ is updated efficiently through $\beta(t) = \sum_{k=1}^{K_B}B_k^* \psi_k(t)$ for any $t \in \mathcal{T}$.

Next, we sample the intercept $\mu$ and the regression coefficients $\bm \alpha$ in \eqref{sofr}--\eqref{sofr_simple}. For simplicity, we omit the intercept and assume that it is already contained in $\bm \alpha$ with the corresponding flat prior (i.e., $\sigma_0 \to \infty$). The full conditional distribution is  $[\bm \alpha \mid \cdots]\sim \mathcal{N}(\bm Q_{\bm \alpha}^{-1} \bm l_{\bm \alpha},\bm Q_{\bm \alpha}^{-1} )$  with 
\begin{align}
\bm Q_{\bm \alpha} = \sigma^{-2} \bm Z'\bm Z + \bm \Sigma_{\bm \alpha}^{-1}  , \quad \bm l_{\bm \alpha} = \sigma^{-2} \bm Z' \bm y_{c'}
\end{align}
where $\bm \Sigma_{\bm \alpha} \coloneqq $ diag$\{\sigma_j^2\}$ is the diagonal matrix of the prior variances on $\bm \alpha$ and $\bm y_{c'} \coloneqq (y_1- \bm X_1^{**}\bm B^*, y_2-\bm X_2^{**}\bm B^*,\ldots, y_n-\bm X_n^{**}\bm B^*)^{'}$ is the $n$-dimensional vector of centered observations. 

The local scale parameters $\{\lambda_k\}_{k=2}^{K-1}$ are updated via the log-volatilities $h_k = \log \lambda_k^2$. Within the Gibbs sampler, the likelihood for this term is given by the second-differenced basis coefficients, $\Delta^2 B_k^* \mid h_k \stackrel{\text{indep}}{\sim}\mathcal{N}(0,\exp(h_k))$ for $k=2,..,K_B-1$. Combining this likelihood with the autoregressive $Z$-distribution model, we note that the dynamic shrinkage prior sampling steps from \cite{kowal2019dynamic} are directly applicable using $\{\Delta^2 B_k^*\}$ as inputs. Specifically, \cite{kowal2019dynamic} obtains a \emph{conditionally} Gaussian likelihood \emph{and} autoregressive model using two parameter expansions: (i) a discrete mixture of Gaussian distributions to approximate the observation equation, which is common for Gaussian stochastic volatility models \citep{omori2007stochastic}, and (ii) a Pólya-Gamma parameter expansion \citep{polson2013bayesian} of the $Z$-distribution. As a result, the full conditional distribution of $\{h_k\}_{k=2}^{K-1}$ is Gaussian with a banded (tridiagonal) precision matrix, resulting in a joint sampler for $\{h_k\}_{k=2}^{K-1}$---and equivalently, $\{\lambda_k\}_{k=2}^{K-1}$---that only requires $\mathcal{O}(K_B)$ computational complexity.  The updates for the parameter expansion variables are identical to those in \cite{kowal2019dynamic}. For the local scale parameters on the boundaries, we update 
\begin{align}
    [\lambda_0^{-2}\mid \cdots]\sim \text{Gamma}(0.01 + 1, \ 0.01 + ({B_1^*}^2 + {B_k^*}^2)/2)
\end{align}
and set $\lambda_1= \lambda_{K_B} = \lambda_0$.  

Conditional on the log-volatilities $\bm h$, we sample the autoregressive parameters $\{\mu_h, \phi\}$ exactly as in \cite{kowal2019dynamic}. 

Lastly, the variance components are updated from $[\sigma^{-2}\mid \cdots]\sim \text{Gamma}(0.01 + n/2, 0.01 + \sum_{i=1}^n (y_i - \mu -  \bm z_i' \bm \alpha - \mathbf {X}_i^{**} \bm B^*   )^2/2)$ and $[\sigma_j^{-2}\mid \cdots]\stackrel{ind}{\sim} \text{Gamma} (0.01 + 1/2, 0.01 + \alpha_j^2/2)$.

In our data analysis, we include nonlinear additive terms using two strategies: (i) piecewise continuous linear splines for mother's age and gestational length and (ii) the proposed B-spline basis expansion with DHS priors on the second-differenced basis coefficients for age-within-cohort. For mother's age and gestational length, the above sampling algorithm for $\bm \alpha$ still applies, but requires careful definition of the $\bm z_i$ components. We construct the piecewise continuous linear splines using knots at ages 18, 24, 29, and 34 for mother's age and weeks 34, 37, 39, and 41 for gestational length. Specifically, we augment $\bm z_i$ with the continuous variables  \texttt{mAge}, $(\texttt{mAge}-18)_{+}$, $(\texttt{mAge}-24)_{+}$, $(\texttt{mAge}-29)_{+}$, $(\texttt{mAge}-34)_{+}$, $\texttt{Gestation}$, $(\texttt{Gestation} - 34*7)_+$,  $(\texttt{Gestation} - 37*7)_+$,  $(\texttt{Gestation} - 39*7)_+$, and $(\texttt{Gestation} - 41*7)_+$, where $(x)_+ = x$ for $\forall x > 0$ and $(x)_+ = 0$ for $\forall x \leq 0$. For age-within-cohort, the model for the regression function features the exact same model specification as for $\beta$ in \eqref{sofr}, so the above blocks for sampling the basis coefficients, log-volatilities, and accompanying autoregressive parameters apply with minor modifications.

\section{Additional Simulation Results}\label{add_sims}

In the main paper, we present simulations to evaluate the proposed BASOFR method and its competitors for point estimation and uncertainty quantification using functional covariates with \textit{seasonality patterns}. In this section, we present simulation results under similar settings but using functional covariates \textit {without seasonality patterns}. Specifically, we keep the simulation settings the same as they are in Section 4.1 of the main paper, but instead set the mean function of the functional covariate $\{X_i(t)\}$ to be $\mu_i(t) = 0 $. The resulting curves $\{X_i(t)\}$ are still smooth, but are no longer seasonal.

\begin{figure}
    \centering
     \includegraphics[width=\textwidth]{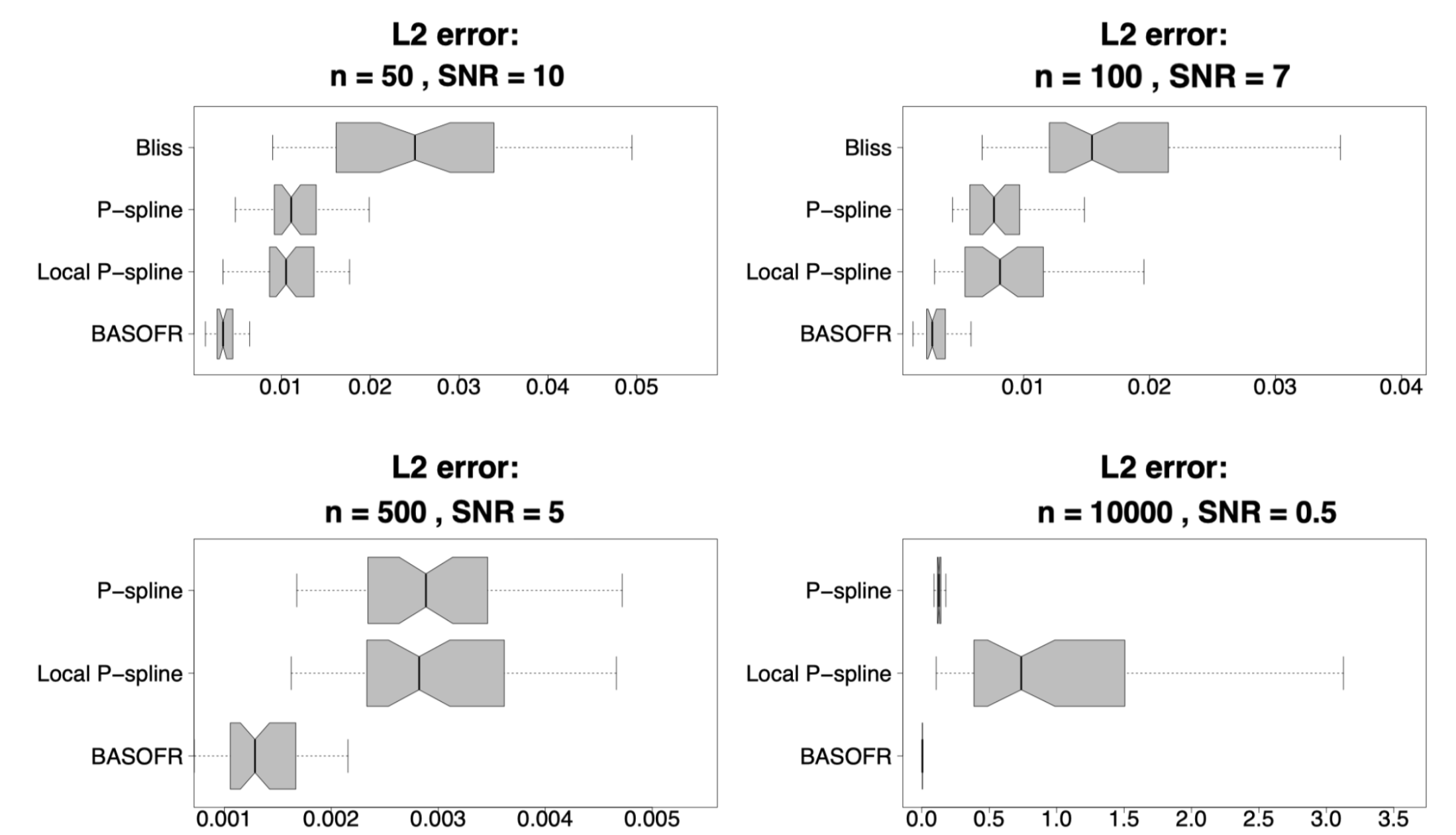}
    \caption{$L_2$-error for  estimating the true regression function using functional covariates without seasonal pattern.}
    \label{fig:l2error_sup}
\end{figure}

\begin{figure}
    \centering
      \includegraphics[width=\textwidth]{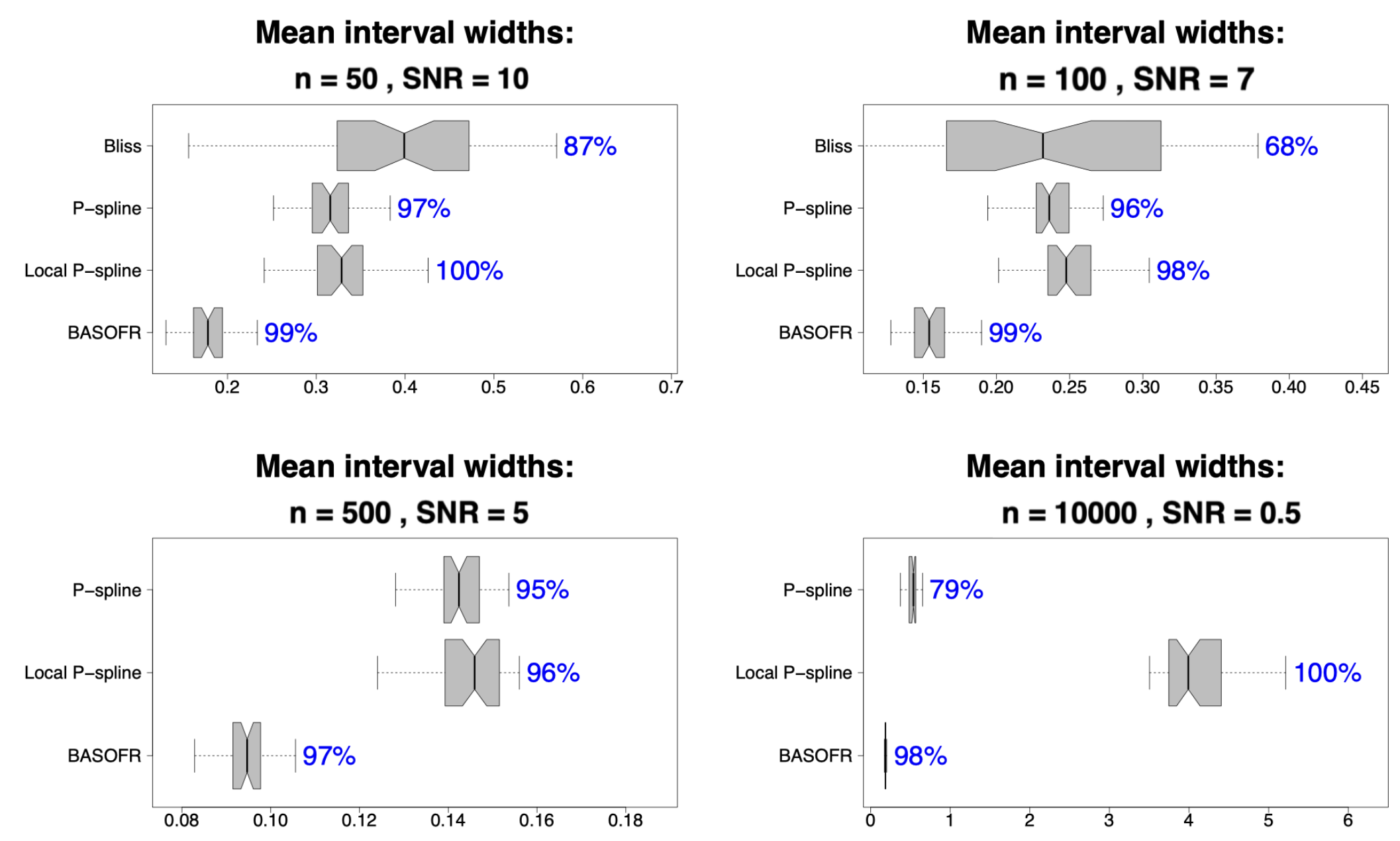} 
    \caption{Mean credible interval widths (boxplots) and empirical pointwise coverage  (blue  annotations) for the 95\% credible intervals computed under each model. The proposed approach (BASOFR) offers substantially more precise uncertainty quantification that maintains or nearly maintains the nominal coverage.}
    \label{fig:uq_sup}
\end{figure}

We present the point estimation and uncertainty quantification results in Figures \ref{fig:l2error_sup} and \ref{fig:uq_sup}, respectively. Notably, the removal of the seasonality component for $\{X_i(t)\}$ makes accurate estimation and inference much easier for all models. Besides that, the results are consistent with those in the main paper in the sense that the proposed BASOFR method produces substantial improvement over competing methods for both point estimation accuracy and uncertainty quantification, especially under larger sample sizes.

\section{Prenatal $\mbox{PM}_{2.5}$ exposures and educational outcomes}\label{add_apps}

In this section, we present additional details and analysis of the North Carolina (NC) education dataset.
%information of the dataset and additional results  of the application study.
\subsection{Autocorrelation among the control variables}

\begin{figure}
    \centering
    \includegraphics[width=\textwidth]{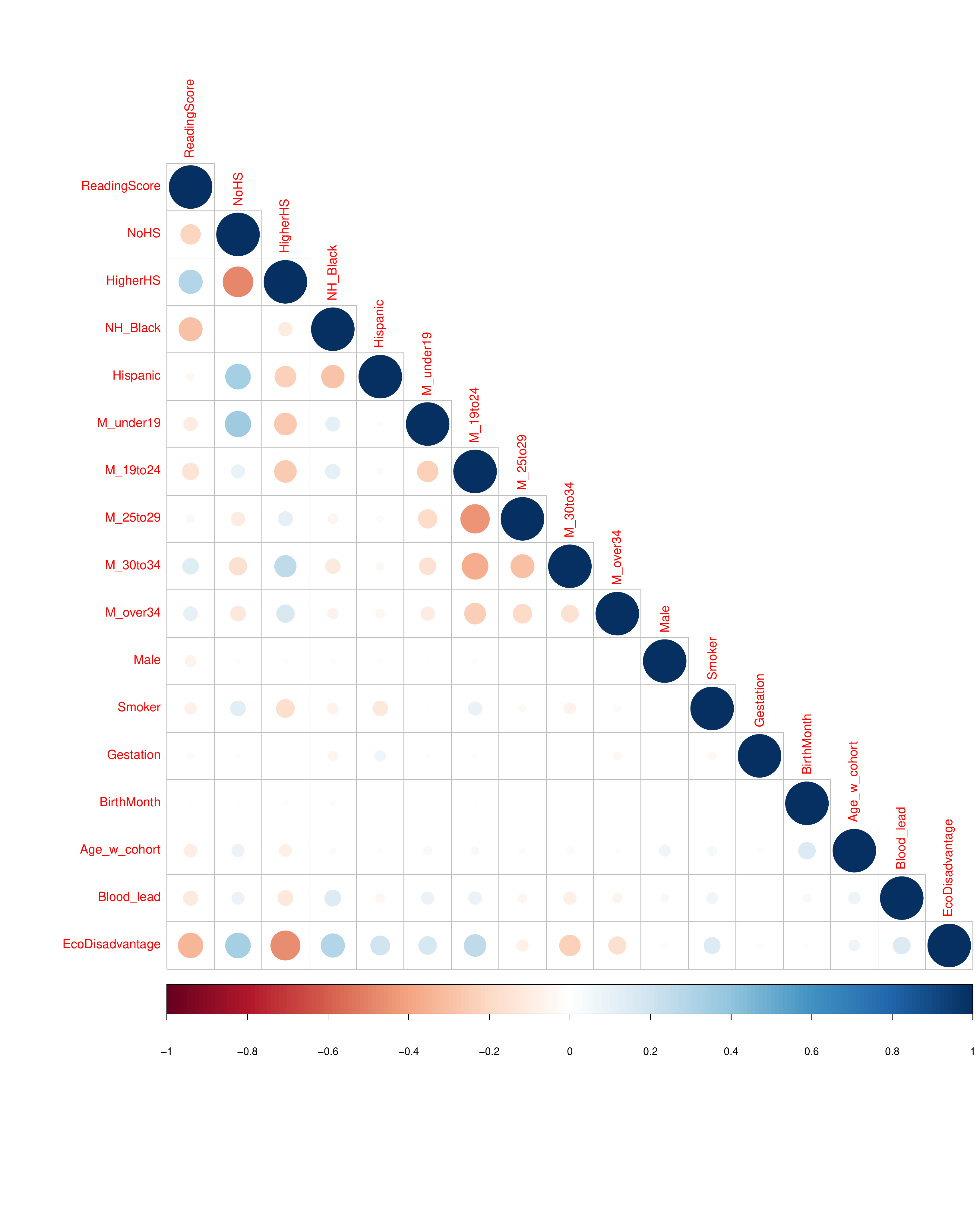}
    \caption{Correlation matrix of the scalar covariates we used in the application study. The \texttt{mAge} (mother's age) variable shows up in the form of five indicator variables: M\_under19, M\_19to24, M\_25to29, M\_30to34, and M\_over34 which indicate the age groups of the mothers}
    \label{fig:corr_scalar}
\end{figure}

We present the correlation matrix of the scalar covariates included in the BASOFR model in Figure~\ref{fig:corr_scalar}.  This figure also includes correlations for specific mother's age groups as these correlation relationships become important for interpreting the result of the application study. Specifically, we found that there are strong correlations between mother's age below 24 and a mother not having a high school degree. This piece of information was used to interpret the regression results regarding the mother's age variable in Figure 11 of the main paper.

\subsection{Details and summary statistics for age-within-cohort and birth month}
We first show how the age-within-cohort (\texttt{Age\_w\_cohort}) variable is computed, followed by further discussion of the results related to this variable. This variable is important: we expect that older students within a cohort may be more intellectually (and emotionally) mature, and thus may perform better on their end-of-grade (EOG) standardized test scores. However, students that are much older than their cohort were likely held back in school or started school later, perhaps due to development difficulties. Thus, we introduce the \texttt{Age\_w\_cohort} variable in the model to capture the effect of being relatively younger or older within the cohort on the testing performance.

This \texttt{Age\_w\_cohort} variable is obtained as follows: we first collect the youngest age requirement for children who entered kindergarten at years 2008, 2009, and 2010 (see Table~\ref{cutoff_age}) . We note that students who entered kindergarten at 2008, 2009, and 2010 would take their EOG test in year 2013, 2014, and 2015, respectively (if they did not skip or repeat a year), so the birth dates in Table~\ref{cutoff_age} can be used to mark the youngest students among their cohorts. For each cohort, we calculate the \texttt{Age\_w\_cohort} variable by taking the differences between each student's date of birth and the youngest student's date of birth (in days). 

\begin{figure}
    \centering
    \includegraphics[width=\textwidth]{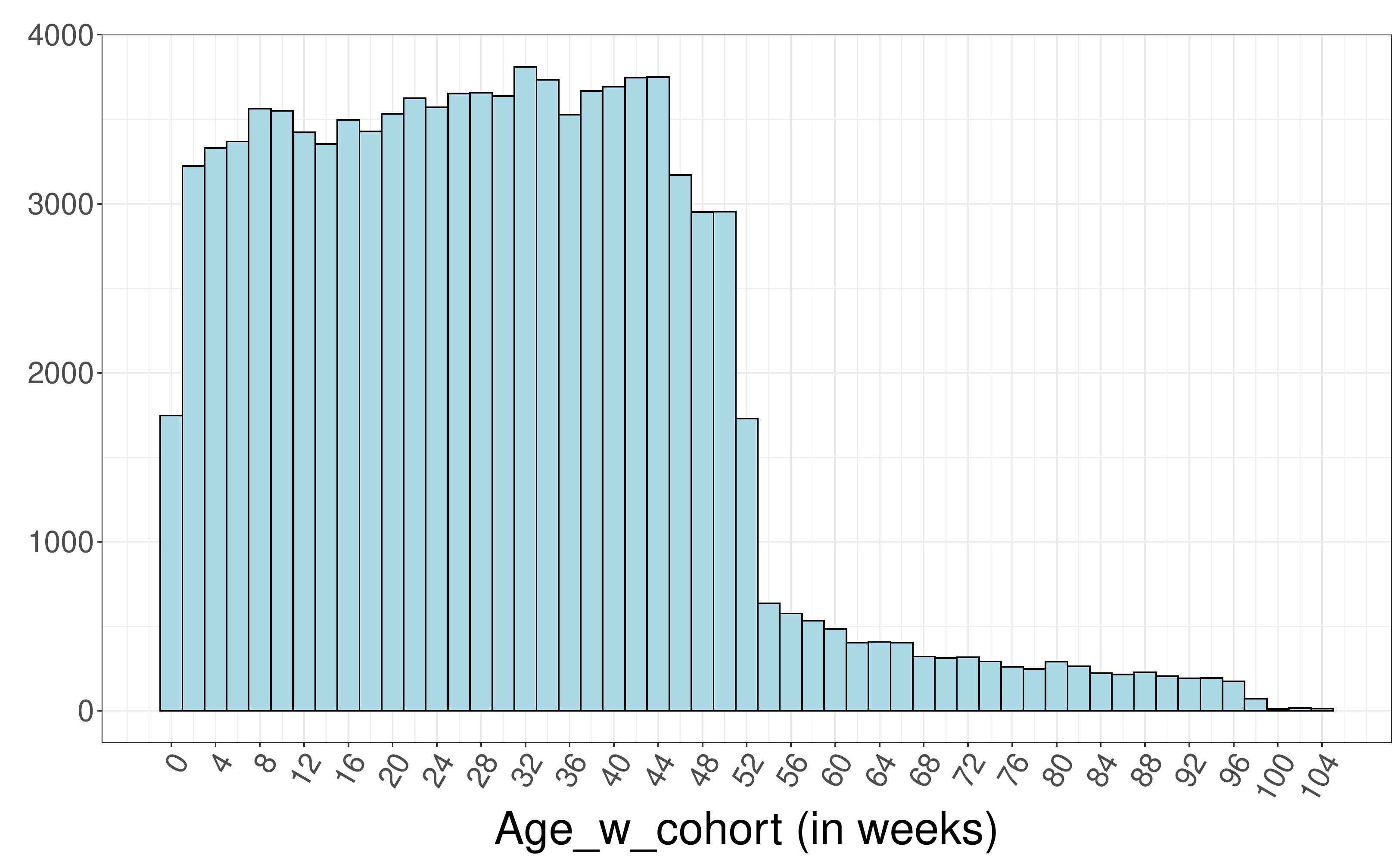}
    \caption{Histogram of the \texttt{Age\_w\_cohort} variable. Data are restricted to $\texttt{Age\_w\_cohort} \in [0, 104]$ weeks, so age differences within a cohort are at most two years.} 
    \label{Age_w_cohort_dist}
\end{figure}

The distribution of the \texttt{Age\_w\_cohort} variable is displayed in Figure~\ref{Age_w_cohort_dist}. We found that most of the students are less than one year older than the youngest student in their cohort  (i.e.,  $\texttt{Age\_w\_cohort} < 365$ days for 92.3\% of students). This means that students who are born in months that are close to and no later than September are very likely to be among the youngest in their  cohorts. These facts should be considered when interpreting the regression results (i.e., Figure 11 in the main paper) for the \texttt{Age\_w\_cohort} and the \texttt{BirthMonth} variable, since these variables have significant overlap. %: the negative estimate for birth months in May-September might actually be capturing the negative impact of being younger than other cohorts. 

\begin{table}
\centering 
\begin{tabular}{rr}
\hline
2008 & October 16, 2003\\ 
2009 & August 16, 2004 \\ 
2010 & Augest 16, 2005 \\ 
\hline
\end{tabular}
\caption{The latest dates of birth for children who entered kindergarten at years 2008, 2009, and 2010, which is used to define the youngest student in each cohort. \label{cutoff_age}}
\end{table}

\begin{figure}
    \centering
    \includegraphics[width=\textwidth]{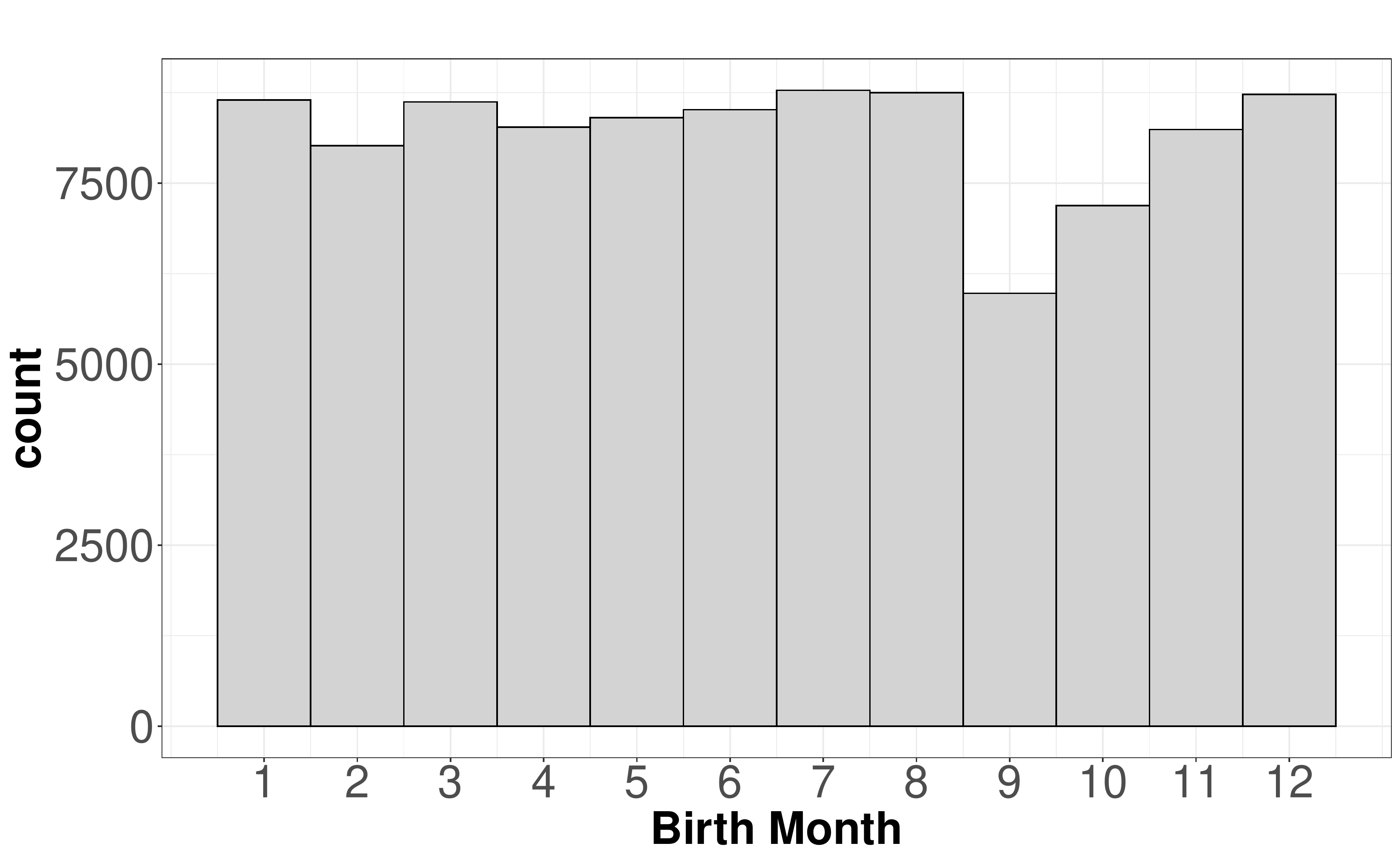}
    \caption{Histogram of the birth month among all students.} 
    \label{BirthMonthDist}
\end{figure}

Lastly, we present the birth month distribution in Figure~\ref{BirthMonthDist}, which shows that the birth months of all students are approximately evenly distributed throughout the year. %This information is relevant to our data analysis because the students with estimated positive effects for $\mbox{PM}_{2.5}$ were almost exclusively born in  October-December. 

%important for our analysis, as it suggests that we need to explain why the birth months of students who receive a positive effect from prenatal $\mbox{PM}_{2.5}$ are concentrated around October to December.

\clearpage

\bibliographystyle{chicago}
\bibliography{main.bib}

\end{document}